\documentclass[useAMS,usenatbib]{mn2e}

\usepackage{graphics}
\usepackage{epsfig}
\usepackage{latexsym}

\usepackage{amsfonts}
\usepackage{amsmath}
\usepackage{amssymb}
\usepackage[english]{babel}
\usepackage{graphicx}
\usepackage{url}
\usepackage{textcomp}
\usepackage{tabularx}

\usepackage{amssymb}
\usepackage{graphics}
\usepackage{graphicx}
\usepackage{epstopdf}

\usepackage[usenames]{color}

\newcommand{\be}{\begin{equation}}
\newcommand{\ee}{\end{equation}}
\newcommand{\ba}{\begin{eqnarray}}
\newcommand{\ea}{\end{eqnarray}}
\newcommand{\bc}{\begin{center}}
\newcommand{\ec}{\end{center}}

\def\mnras{{MNRAS}}
\def\apJ{{ApJ}}
\def\aJ{{AJ}}
\def\apJL{{ApJL}}
\def\aap{{A\&A}}

\def\aapr{A\&ARv}
\def\nat{Nature}
\def\pasp{PASP}

\def\aas{A\&AS}

\title[SN~2013df]{SN~2013df, a double-peaked IIb supernova from a compact progenitor and an extended H envelope}
\author[Morales-Garoffolo et al.]{A. Morales-Garoffolo$^{1}$\thanks{amg@ice.csic.es}, N. Elias-Rosa$^{2}$, S. Benetti$^{2}$, S. Taubenberger$^{3}$, \newauthor E. Cappellaro$^{2}$, A. Pastorello$^{2}$, M. Klauser$^{3}$,  S. Valenti$^{4}$$^{,}$$^{5}$, S. Howerton$^{6}$,\newauthor P. Ochner$^{2}$, N. Schramm$^{7}$, A. Siviero$^{8}$, L. Tartaglia$^{2}$$^{,}$$^{8}$, L. Tomasella$^{2}$  \\ 
$^1$Institut de Ci\`encies de l'Espai (CSIC-IEEC), Campus UAB,  Torre C5, 2a planta, 08193 Barcelona, Spain \\
$^2$INAF Osservatorio Astronomico di Padova, Vicolo dell'Osservatorio 5, 35122 Padova, Italy\\
$^{3}$Max-Planck-Institut f\"{u}r Astrophysik, Karl-Schwarzschild-Str. 1, 85741 Garching bei M\"{u}nchen, Germany\\
$^{4}$Las Cumbres Observatory Global Telescope Network, 6740 Cortona Dr., Suite 102, Goleta, CA 93117, USA\\
$^{5}$Department of Physics, University of California, Santa Barbara, Broida Hall, Mail Code 9530, Santa Barbara, CA 93196-9530, USA\\
$^{6}$1401 South A, Arkansas City, KS 67005, USA\\
$^{7}$384 Janes Chapel Road, Oxford, NJ 07863, USA\\
$^{8}$Dipartimento di Fisica e Astronomia Galileo Galilei, Universit\'a di Padova, Vicolo dell'Osservatorio 3, 35122 Padova, Italy\\}

\begin{document}

\date{}

\pagerange{\pageref{firstpage}--\pageref{lastpage}} 
\pubyear{}

\maketitle

\label{firstpage}

\begin{abstract}
Optical observations of the type IIb SN~2013df from a few days to about 250 days after explosion are presented. These observations are complemented with UV photometry taken by \textit{SWIFT} up to 60 days post-explosion. The double-peak optical light curve is similar to those of SNe 1993J and 2011fu although with different decline and rise rates. From the modelling of the bolometric light curve, we have estimated that the total mass of synthesised $^{56}$Ni in the explosion is $\sim0.1$ M$_{\odot}$, while the ejecta mass is $0.8\textendash1.4$ M$_{\odot}$ and the explosion energy $0.4\textendash1.2 \times 10^{51}$erg. In addition, we have estimated a lower limit to the progenitor radius ranging from $64-169$ $R_{\odot}
$. The spectral evolution indicates that SN~2013df had a hydrogen envelope similar to SN~1993J ($\sim 0.2$ M$_{\odot}$). 
The line profiles in
nebular spectra  suggest 
that the explosion was asymmetric with the presence of clumps in the ejecta, while the [O\,{\sc i}] $\lambda$$\lambda$$6300$, $6364$ luminosities, may indicate that the progenitor of SN~2013df was a relatively low mass star ( $\sim 12-13 $ M$_{\odot}$). 
\end{abstract}

\section{Introduction}

Massive stars  \citep[$M_{\rm ZAMS} > 8$ M$_{\odot}$; see e.g.][]{heger03} usually end their lives exploding as Core-Collapse Supernovae (CC-SNe) leaving behind a compact remnant (neutron star or black hole depending on the mass of the star that exploded). The stellar configuration prior to explosion is what leads to the different observational sub types of CC-SNe. Specifically, type IIb SNe are thought to arise from the explosion of stars that had retained a small part (a few tenths of M$_{\odot}$) of their hydrogen layer, and along with type Ib and Ic SNe belong to the so-called stripped-envelope SNe category (SE-SNe) \citep{clocchiatti96}. The reason why type IIb SN progenitors have lost most of their hydrogen envelope prior to exploding is unclear, although two scenarios are contemplated: 1) Transfer of most of the stellar envelope of the progenitor to a companion star \citep[see e.g.,][]{binaryIIb}; 2) Mass loss due to stellar winds \citep{smow06,Puls08}. 
 Type IIb SNe are relatively infrequent events. Their rate among a volume limited sample of 81 type II SNe was estimated by \cite{Li11} to be $11.9\% ^{+3.9}_{-3.6}$. The Asiago SN catalogue lists about 86 type IIb SNe\footnote[9]{http://sngroup.oapd.inaf.it} but only a few of them have been extensively monitored. Their spectra show a 
transition from being dominated by 
hydrogen at early phases, to predominance of He\,{\sc i} features at later times \citep{Fpk88}. Concerning their light curves (LCs), some type IIb SNe present clear double-peaked LCs in all optical bands, as in the case of SN~1993J \citep[e.g.,][]{Richmond94} and SN~2011fu \citep{11fu}. Other IIb SNe show the first peak only in some of their optical and UV bands, e.g. SNe 2008ax \citep{Roming09} and 2011dh \citep{Arcavi11dh}. This is probably due to the short duration  of the first peak in these SNe that can be missed by observations. The first peak of the LCs of IIb SNe is attributed to shock wave heating of the hydrogen envelope, while the longer lasting secondary peak is powered by the decay of $^{56}$Ni \citep{Ws94}. 
Other SNe, that are not of type IIb, have also presented an initial peak in their optical LCs before climbing to a secondary maximum. This is the case for example of type II-peculiar 1987A \citep[e.g.,][]{Hamuy88}, or type Ib/Ic SN 2005bf \citep{Folatelli06}. For SN 1987A, the initial peak has also been explained in terms of an adiabatic cooling of its stellar envelope after the shock breakout that follows core-collapse of massive stars. However, for SN 2005bf \citep{Folatelli06} gave an alternative explanation for the early peak calling for a double-peaked  $^{56}$Ni distribution, this is, some $^{56}$Ni is mixed in the outer layers of the ejecta.  

Confirmed 
progenitor 
identifications for type IIb SNe have been achieved for SN~1993J  \citep{
Al94,Mnd09}  and  
for SN~2011dh \citep{Mnd11,VD11,VD13}. In both cases the progenitor stars were revealed to be yellow supergiants. For SN~2008ax, the progenitor was also found in archival images \citep{Li08,Crockett08}, but there were multiple configurations that matched its observed colours. Two scenarios were proposed, one consisting of a single progenitor and another based on a binary system, but in both cases the explosion involved a supergiant. With respect to the binary scenario, a blue companion to SN~1993J was identified about ten years after its explosion \citep{Mnd04}, and possibly the companion of SN~2001ig was also found by \cite{Ryder2006}. Recently, the indirect detection of the  Wolf Rayet-like progenitor of type IIb SN 2013cu  has been possible from spectral observations of its stellar wind right after explosion \citep{Gal-Yam14}.\\ The most recent probable identification of a type IIb SN progenitor from Hubble Space Telescope (HST) archival images has been that of the yellow 
supergiant precursor of the type IIb SN~2013df \citep{VD13c}.\\ SN~2013df, having coordinates $\alpha =12^{\rmn{h}} 26^{\rmn{m}} 
29\fs33 $ and $\delta = +31\degr 13\arcmin 38.\arcsec3$ (J2000), was discovered in the galaxy NGC~4414 by F. Ciabattari and E. Mazzoni of the Italian Supernovae Project (ISSP)\footnote[10]{http://italiansupernovae.org/en.html}, on 2013 June  7.87 UT \citep{cbet13df}. SN~2013df is the second SN discovered in NGC~4414, the first being the type Ia SN 1974G. A spectrum of SN~2013df taken soon after discovery (2013 June 10.8 UT) showed characteristics of a type II supernova resembling early spectra of the type IIb SN~1993J \citep{cbet13df}. Soon after discovery of the SN, X-ray detection coming from a source  
at the SN position (2.1 arcsec offset) was reported by \citet{Li13}. The absorption corrected average luminosity ($0.3-10 $\,keV) of their observations, which spanned from 2013 June 13 to 2013 June 19, was $7\times10^{39}$erg~s$^{-1}$. This luminosity most likely has its origin in the SN, as it falls in the range of other X-ray detected SNe, and it 
is less probable to arise from diffuse emission of NGC~4414. Some early time optical data of SN~2013df were presented in \cite{VD13c} and confirmed its type IIb nature. In this manuscript we present the analysis of additional optical photometric and spectroscopic data for SN~2013df, as well as ultraviolet and late time data. The monitoring of SN~2013df from early to late time after explosion, has offered us a unique opportunity to investigate the evolution of the observed properties of a relatively nearby type IIb SN, which presents double-peaked LCs in all optical bands, and for which the plausible progenitor has been identified. In section 2, we summarize the reduction and calibration process of our observational data of SN~2013df. In section 3 we present our photometric results. The spectroscopic analysis is described in section 4. In section 5 we derive some constraints on the progenitor's characteristics, and in section 6 we summarize our results. In Appendix \ref{appendix}, we provide details of the 
instrumental set-ups used in the acquisition of our data of SN~2013df.

\section{Observational data}
\subsection{Distance, reddening and host galaxy}

Throughout this work, we will adopt as the distance modulus to NGC~4414 $\mu = 31.65 \pm 0.30$ mag. This estimation was obtained as the weighted mean value of distance moduli provided by the NASA/IPAC extragalactic database (NED). The redshift for SN~2013df is assumed to be that of its host galaxy as given by NED ($v_{\rm recession}= 716\pm 6$ $\rm km\,s^{-1}$). The reddening in the line of sight of the SN due to the Milky Way  \citep{Schfink} is $E(B-V)_{{\rm MW}}=0.017\pm0.001$~mag. Thus, adopting  $E(B-V)_{{\rm NGC\;4414}}=0.081\pm0.016$~mag as the SN reddening in the host-galaxy \citep{VD13c}, we assume in the rest of this manuscript $E(B-V)_{{\rm Total}}=0.098\pm0.016$~mag as the total reddening toward SN~2013df.\\ \\ \cite{tomasella14} performed a study of the distribution of the subtypes of 78 CC-SNe in dwarf compared to giant galaxies. In their sample there were four type IIb SNe whose hosts 
were luminous 
galaxies. Keeping in mind the very small statistics and the uncertainties in the early classification of type IIb SNe, their result does not confirm the excess of type IIb SNe in dwarf hosts  claimed by \cite{Arcavi2010} and, on the contrary, it is in agreement with \cite{Sanders12}, who found no statistical differences between the metallicity distributions of type Ib and Ic or between type Ib and IIb SNe.
With our adopted distance, the reddening towards 
NGC~4414 due to 
the 
Milky Way, and the apparent \textit{V} magnitude given by SIMBAD\footnote [11]{http://simbad.u-
strasbg.fr/simbad/} we obtain for NGC~4414 $ V = -21.6$. 
So, NGC~4414 is another example of a luminous (giant) host for a type IIb SN.	

\subsection{Ground-based optical photometric and spectroscopic data: reduction and calibration process.}

We started an optical observational follow-up campaign of SN~2013df on 2013 June 11.93 UT (a few days after its discovery), the campaign was suspended at the end of July 2013 when the supernova went behind the Sun and resumed again to acquire some late time data from November 20th, 2013 to February 11th, 2014. Our data were obtained at different sites (see Appendix\,\ref{appendix}). The reduction and calibration processes of the data are described below:

\begin{enumerate}
 \item{Photometry}
         
         Non-amateur optical photometric data were reduced (overscan, bias and flat field corrected) within the {\sc iraf}\footnote[12]{Image reduction and Analysis Facility, a software system distributed by the National Optical Astronomy Observatories (NOAO)} environment  for all instruments except for data obtained at LT, for which reductions are usually performed with specific pipelines.  The measurement of instrumental magnitudes of the SN were performed via Point Spread Function (PSF) fitting by means of SNOoPY\footnote[13]{Cappellaro 2014 SNOopy: a package for SN photometry, http://sngroup.oapd.inaf.it/snoopy.html}. A total of 5 local stars, identified in Figure \ref{fig:13dfsequence}, were chosen to derive the instrumental PSF and to calibrate the optical \textit{UBVRI} SN LCs.

         Once instrumental magnitudes were measured, calibration of the 
local stellar field to a standard photometric system  was performed by means of the zero points and colour terms derived thanks to the observation of standard Landolt fields \citep{Landolt} on photometric nights. The apparent magnitudes of the stellar sequence and their associated errors are reported in Table~\ref{sequence_mags}  both in the  Johnson Cousins - JC - system and in the Sloan system, and are the mean values of the magnitudes obtained over photometric nights while the errors are the r.m.s. uncertainties over these nights. The SN magnitudes were derived relative to the mean magnitudes of our stellar sequence.\\ The unfiltered magnitudes obtained from the images provided by F. Ciabattari, K. Itagaki and S. Howerton were rescaled to JC \textit{V} band or \textit{R} band, depending on the wavelength peak efficiency of the detectors. 
The data provided by N. Schramm were taken with a colour camera, which has a colour filter array (or Bayer-mask) that arranges the red, green, and blue response on each pixel of the CCD. We subdivided in blue, green, and red images each of the frames. One of the main problems when doing this is that gaps result between the pixels of the images of the same colour. After re-binning the images, we derived SN 
magnitudes from these data and we rescaled to \textit{B}, \textit{V}, and \textit{R} band magnitudes those derived from the blue, green and red images respectively. 

We note that we have not performed S-corrections \citep{stritz02,pignata04} to our photometric data of SN~2013df.\\

\begin{table*}
\centering
\caption{Magnitudes and associated errors in the Johnson-Cousins and Sloan systems of the stellar sequence used in the calibration process of SN~2013df's photometry.}
\begin{tabular}{ccccccccc} \hline 
    Star    &    \textit{U}  &  \textit{B} & \textit{V}                & \textit{R} & \textit{I} &    \textit{u}   & \textit{r}                & \textit{i} \\ 
            & (mag) & (mag) & (mag) & (mag)& (mag) & (mag) & (mag)& (mag) \\
  \hline
1  &$     $ & $17.36\pm 0.03$      & $15.83\pm0.02$ & $14.73\pm0.02$  & $13.28\pm0.02$            &$19.49\pm0.00  $*& $15.24\pm0.02 $         & $14.04\pm0.06$ \\ 
2  &$16.41\pm 0.04 $ & $16.40\pm 0.04$      & $15.78\pm0.03$ & $15.33\pm0.02$  & $14.97\pm0.02$ &$17.43\pm0.06  $& $15.68\pm0.01 $         & $15.50\pm0.01$  \\ 
3  &$16.47\pm 0.04 $ & $15.88\pm 0.03$      & $14.87\pm0.02$ & $14.29\pm0.03$  & $13.74\pm0.02$ &$17.32\pm0.03  $& $14.59\pm0.01 $         & $14.28\pm0.03$       \\ 
4  &$16.55 \pm 0.04 $ & $16.62\pm 0.05$      & $16.01\pm0.03$ & $15.70\pm0.03$  & $15.31\pm0.02$ & & &  \\ 
5  &$     $ & $17.74\pm 0.04$      & $16.29\pm0.03$ & $15.48\pm0.02$  & $14.71\pm0.02$ & & &                 \\

\hline
\end{tabular} \\
* The star was only measured in one epoch.
\label{sequence_mags}
\end{table*}

 \begin{figure}
 \includegraphics[width=8.6cm]{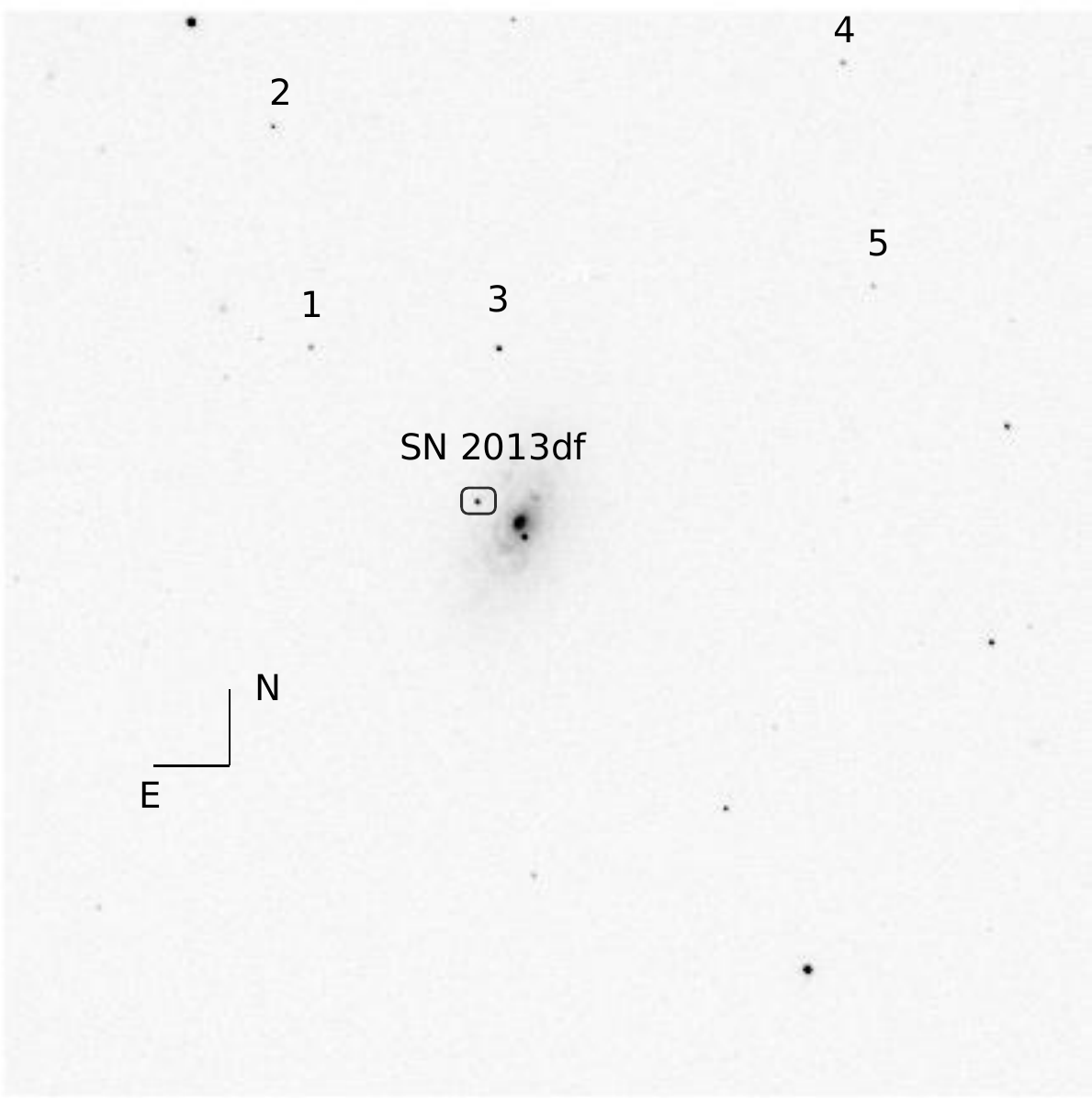}
 \caption{\textit{V} band image of NGC~4414 taken with TJO+MEIA on June 20th 2013. The stars used for the photometric calibration of SN~2013df are labelled, and the field of view is approximately 12'.3 x 12'.3. }
 \label{fig:13dfsequence}
\end{figure}

 \item{Spectroscopy}
 
 Optical spectroscopic data were reduced within {\sc iraf} (overscan, bias and flat fielding). The spectra were extracted, wavelength calibrated by means of the spectra of arc lamps taken in the same configuration as the spectra of the SN, and flux calibrated by means of the use of sensitivity functions obtained after observations of spectroscopic standards. Atmospheric corrections were also performed  dividing by spectra of telluric lines obtained from the spectroscopic standards. When spectra were obtained with different grisms, they were scaled and combined to obtain final spectra. The spectra wavelength and flux calibration were checked using night sky lines and the photometry of the SN at the nearest epoch respectively.\\ 
 \end{enumerate}

 \subsection{Space-based \textit{SWIFT} optical and ultraviolet data: reduction and calibration process.}

  Our ground-based photometry was complemented with data taken by UVOT on board the \textit{SWIFT} satellite (see Appendix \ref{appendix}). 
  The calibrated UVOT SN images were retrieved from the \textit{SWIFT} data archive \footnote [14]{https://heasarc.gsfc.nasa.gov/cgi-bin/W3Browse/swift.pl}. Estimates of the SN magnitudes were obtained using the task {\em uvotsource}  included in the {\em ftools} packages distributed by NASA's HEASARC. The task performs aperture photometry, corrected for the detector coincidence losses, integrating the flux in a user defined aperture. In the particular case, to reduce the contamination by galaxy background, the SN magnitude was measured with a 3 arcsec aperture. We also measured, as reference, the magnitude of field stars with the same 3" aperture along with a 5" aperture.  The reference stars were used to derive an aperture correction and, by including  some of the local standards, were used to cross-check the photometric calibration for the optical bands obtained at other telescopes. The \textit{SWIFT} UV photometry for SN~2013df is reported in Table~\ref{uv_phot}.

\section{Photometric Results}
\subsection{UV-Optical Light Curves}
\label{subsection:UVopt}

\begin{figure*}
 \includegraphics[trim=0cm 0cm 0cm 6cm,width=15.6cm]{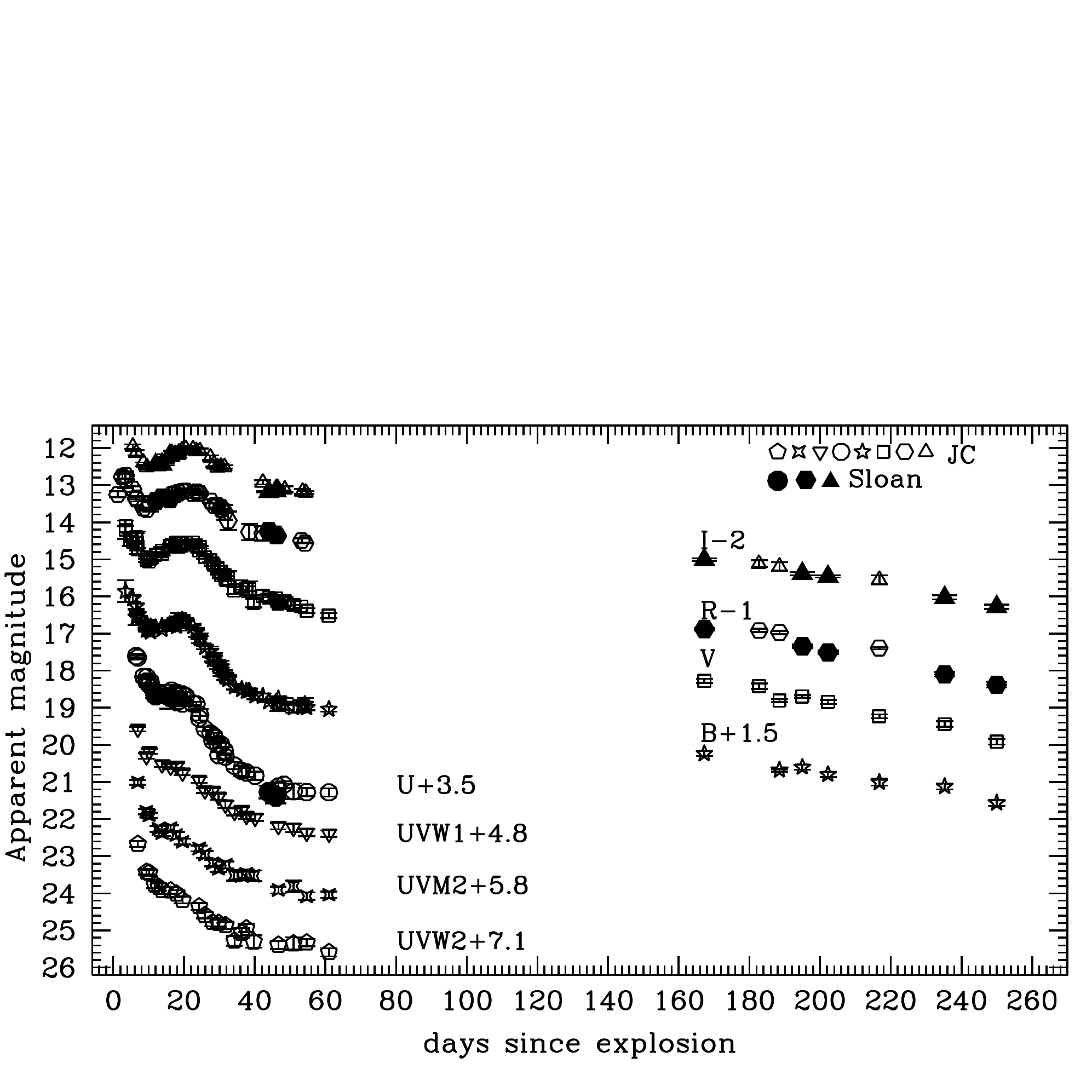}
 \caption{UV-optical light curves of SN~2013df. The \textit{uri} Sloan data have also been included in these light curves using the approximations to the JC system given in \citet{Jordi}. The transformed Sloan data points are depicted in the figure with filled symbols. The assumed explosion epoch is $\rm JD = 2456450.0\pm0.9$. The LCs have been shifted for clarity by the values indicated in the figure.}
 \label{fig:apparentLCs}
\end{figure*}

The optical \textit{UBVRI} and \textit{SWIFT} \textit{UVW2}, \textit{UVM2}, \textit{UVW1} light curves of SN~2013df spanning from discovery up to approximately 250 days post discovery is shown in Figure \ref{fig:apparentLCs}. The SN's optical apparent magnitudes together with the errors are reported in Table~\ref{UBVRI_phot}. We have included in our optical light curves some filtered and unfiltered amateur data. Note that \textit{uri} Sloan data obtained at the LT (presented in Table~\ref{uri_phot}), have been approximated to the JC \textit{URI} using the 
relations in \citet{Jordi} and 
then added to the \textit{UBVRI} 
light curves of Figure \ref{fig:apparentLCs} using filled symbols. The transformations in \citet{Jordi}  are colour transformations between \textit{ugriz} Sloan Digital Sky Survey (SDSS) photometry and JC photometry. When applied to our stellar sequence, we found a fair agreement with our own JC photometry (differences ranging $\sim$ 0.1-0.3 mag). The transformations seem to work fairly well for our SN photometry, specially at early phases, although they still should be considered approximations.\\

\begin{table*}
\centering
\caption{\textit{SWIFT} UV photometry of SN~2013df taken with UVOT on board \textit{SWIFT} + MIC.}
\begin{tabular}{ccccccc} \hline 
    Date & JD & Phase$^{a}$    &    \textit{UVW2}  &  \textit{UVM2} & \textit{UVW1}   & \\ 
         & (+2400000.00) & (d) & (mag)& (mag)&(mag)&\\
  \hline
2013/06/13	&	56456.90	&	6.9	 &$   15.57	\pm  0.08  $&$  15.21	\pm	0.06	$&$   14.77	\pm	0.06        $	\\
2013/06/15	&	56459.35	&	9.4	 &$   16.32	\pm  0.08  $&$  16.00	\pm	0.07	$&$   15.50	\pm	0.07        $	\\
2013/06/16	&	56460.17	&	10.2	 &$  16.35	\pm  0.08  $&$  16.10	\pm	0.07	$&$   15.37	\pm	0.07        $	\\
2013/06/18	&	56461.75	&	11.7	 &$  16.67	\pm  0.09  $&$                   	$&$                                 $	\\
2013/06/19	&	56462.82	&	12.8	 &$  16.73	\pm  0.10  $&$                   	$&$                                 $	\\
2013/06/20	&	56463.82	&	13.8	 &$  16.81	\pm  0.09  $&$  16.58	\pm	0.07	$&$   15.72	\pm	0.07        $	\\
2013/06/22	&	56466.19	&	16.2	 &$  16.82	\pm  0.09  $&$  16.45	\pm	0.08	$&$   15.79	\pm	0.07        $	\\
2013/06/24	&	56467.97	&	18.0	 &$  16.91	\pm  0.09  $&$  16.63	\pm	0.09	$&$   15.76	\pm	0.07        $	\\
2013/06/26	&	56469.56	&	19.6	 &$  17.08	\pm  0.09  $&$  16.81	\pm	0.08	$&$   15.95	\pm	0.07        $	\\
2013/06/30	&	56474.29	&	24.3	 &$  17.25	\pm  0.09  $&$  16.99	\pm	0.08	$&$   16.15	\pm	0.07        $	\\
2013/07/02	&	56475.91	&	25.9	 &$  17.49	\pm  0.09  $&$  17.15	\pm	0.09	$&$   16.43	\pm	0.07        $	\\
2013/07/04	&	56478.13	&	28.1	 &$  17.68	\pm  0.11  $&$  17.39	\pm	0.09	$&$   16.44	\pm	0.08        $	\\
2013/07/06	&	56479.76	&	29.8	 &$  17.70	\pm  0.10  $&$  17.54	\pm	0.09	$&$   16.60	\pm	0.08        $	\\
2013/07/08	&	56481.76	&	31.8	 &$  17.76	\pm  0.10  $&$  17.43	\pm	0.09	$&$   16.81	\pm	0.09        $	\\
2013/07/10	&	56484.25	&	34.2	 &$  18.18	\pm  0.14  $&$  17.72	\pm	0.14	$&$   17.01	\pm	0.09        $	\\
2013/07/12	&	56486.33	&	36.3	 &$  17.97	\pm  0.13  $&$  17.73	\pm	0.12	$&$   16.96	\pm	0.10        $	\\
2013/07/14	&	56487.64	&	37.6	 &$  17.84	\pm  0.11  $&$  17.67	\pm	0.11	$&$   17.11	\pm	0.09        $	\\
2013/07/16	&	56489.75	&	39.8	 &$  18.20	\pm  0.17  $&$  17.73	\pm	0.14	$&$   17.16	\pm	0.09        $	\\
2013/07/23	&	56496.68	&	46.7	 &$  18.29	\pm  0.11  $&$  18.11	\pm	0.09	$&$   17.39	\pm	0.09        $	\\
2013/07/27	&	56501.03	&	51.0	 &$  18.25	\pm  0.17  $&$  18.01	\pm	0.15	$&$   17.44	\pm	0.11        $	\\
2013/07/31	&	56504.73	&	54.7	 &$  18.23	\pm  0.10  $&$  18.28	\pm	0.10	$&$   17.57	\pm	0.09        $	\\
2013/08/04	&	56509.43	&	59.4	 &$  18.83	\pm  0.39  $&$  18.68	\pm	0.32	$&$   17.39	\pm	0.22        $	\\
2013/08/06	&	56511.01	&	61.0	 &$  18.49	\pm  0.11  $&$  18.24	\pm	0.09	$&$   17.59	\pm	0.09        $	\\

\hline
\end{tabular} \\
\begin{flushleft}
 $^{a}${Phase in days with respect to the adopted explosion date JD = $2456450.0\pm0.9$}\\
 
\label{uv_phot}
\end{flushleft}

\end{table*}

\begin{table*}
\centering
\caption{Optical Johnson Cousins photometry of SN~2013df.}
\begin{tabular}{ccccccccc} \hline 
    Date & JD & Phase$^{a}$    &    \textit{U}  &  \textit{B} & \textit{V}                & \textit{R} & \textit{I} & Instrument $^{b}$\\ 
         & (+2400000.00)   & (d) & (mag)& (mag)&(mag)&(mag)&(mag)& \\
  \hline
2013/06/07	&	56451.37	&	1.4	 &$ 	  	$&$	  	$&$	  	$&$	14.25	\pm	0.08	$&$	  	$&	F.Ciabattari	\\
2013/06/08	&	56452.40	&	2.4	 &$ 	  	$&$	  	$&$	  	$&$	13.77	\pm	0.07	$&$	  	$&	S.Donati	\\
2013/06/09	&	56453.41	&	3.4	 &$ 	  	$&$	  	$&$	14.10	\pm	0.01	$&$	  	$&$	  	$&	S.Howerton T18	\\
2013/06/09	&	56453.41	&	3.4	 &$ 	  	$&$	  	$&$	  	$&$	13.74	\pm	0.08	$&$	  	$&	K.Itagaki	\\
2013/06/10	&	56453.57	&	3.6	 &$ 	  	$&$	14.36	\pm	0.30	$&$	14.20	\pm	0.24	$&$	13.85	\pm	0.23	$&$	  	$&	N.Schramm	\\
2013/06/10	&	56454.70	&	4.7	 &$ 	  	$&$	  	$&$	14.46	\pm	0.18	$&$	  	$&$	  	$&	S.Howerton T5	\\
2013/06/11	&	56455.43	&	5.4	 &$ 	  	$&$	14.55	\pm	0.09	$&$	14.53	\pm	0.02	$&$	14.11	\pm	0.07	$&$	13.96	\pm	0.06	$&	MEIA	\\
2013/06/12	&	56456.42	&	6.4	 &$ 	14.11	\pm	0.06	$&$	14.77	\pm	0.08	$&$	14.55	\pm	0.02	$&$	14.34	\pm	0.04	$&$	14.13	\pm	0.06	$&	MEIA	\\
2013/06/12	&	56456.70	&	6.7	 &$ 	  	$&$	  	$&$	14.55	\pm	0.30	$&$	  	$&$	  	$&	S.Howerton T5	\\
2013/06/13	&	56456.57	&	6.6	 &$ 	  	$&$	15.05	\pm	0.21	$&$	14.40	\pm	0.11	$&$	14.37	\pm	0.06	$&$	  	$&	N.Schramm	\\
2013/06/13	&	56456.90	&	6.9	 &$ 	14.15	\pm	0.05	$&$	14.98	\pm	0.04	$&$	14.73	\pm	0.04	$&$	  	$&$	  	$&	UVOT	\\
2013/06/14	&	56458.43	&	8.4	 &$ 	14.66	\pm	0.03	$&$	15.22	\pm	0.01	$&$	  	$&$	14.62	\pm	0.01	$&$	14.39	\pm	0.01	$&	MEIA	\\
2013/06/15	&	56459.35	&	9.4	 &$ 	14.79	\pm	0.05	$&$	15.40	\pm	0.04	$&$	15.02	\pm	0.05	$&$	  	$&$	  	$&	UVOT	\\
2013/06/15	&	56459.43	&	9.4	 &$ 	14.67	\pm	0.05	$&$	15.24	\pm	0.11	$&$	14.95	\pm	0.03	$&$	14.66	\pm	0.03	$&$	14.51	\pm	0.06	$&	MEIA	\\
2013/06/16	&	56460.42	&	10.4	 &$ 	14.79	\pm	0.06	$&$	15.31	\pm	0.01	$&$	  	$&$	  	$&$	  	$&	MEIA	\\
2013/06/16	&	56460.17	&	10.2	 &$ 	14.88	\pm	0.05	$&$	15.41	\pm	0.05	$&$	15.06	\pm	0.05	$&$	  	$&$	  	$&	UVOT	\\
2013/06/17	&	56460.59	&	10.6	 &$ 	  	$&$	15.45	\pm	0.03	$&$	14.98	\pm	0.11	$&$	14.50	\pm	0.21	$&$	  	$&	N.Schramm	\\

2013/06/18	&	56461.92	&	11.9	 &$ 	15.16	\pm	0.08	$&$	15.34	\pm	0.01	$&$	14.87	\pm	0.01	$&$	14.42	\pm	0.02	$&$	14.46	\pm	0.03	$&	RATCam	\\

2013/06/20	&	56463.82	&	13.8	 &$ 	15.14	\pm	0.05	$&$	15.39	\pm	0.04	$&$	14.85	\pm	0.05	$&$	  	$&$	  	$&	UVOT	\\
2013/06/20	&	56463.88	&	13.9	 &$ 	  	$&$	15.31	\pm	0.02	$&$	14.77	\pm	0.11	$&$	14.34	\pm	0.07	$&$	14.44	\pm	0.03	$&	RATCam	\\
2013/06/22	&	56465.92	&	15.9	 &$ 	15.20	\pm	0.32	$&$	15.25	\pm	0.02	$&$	14.63	\pm	0.01	$&$	14.38	\pm	0.08	$&$	14.21	\pm	0.04	$&	RATCam	\\
2013/06/22	&	56466.19	&	16.2	 &$ 	15.21	\pm	0.06	$&$	15.28	\pm	0.05	$&$	14.65	\pm	0.05	$&$	  	$&$	  	$&	UVOT	\\
2013/06/22	&	56466.42	&	16.4	 &$ 	15.03	\pm	0.07	$&$	15.26	\pm	0.11	$&$	14.61	\pm	0.03	$&$	14.26	\pm	0.04	$&$	14.13	\pm	0.06	$&	MEIA	\\
2013/06/23	&	56467.42	&	17.4	 &$ 	15.13	\pm	0.09	$&$	15.18	\pm	0.11	$&$	14.65	\pm	0.06	$&$	14.24	\pm	0.04	$&$	14.16	\pm	0.07	$&	MEIA	\\
2013/06/24	&	56467.97	&	18.0	 &$ 	15.35	\pm	0.05	$&$	15.31	\pm	0.04	$&$	14.55	\pm	0.06	$&$	  	$&$	  	$&	UVOT	\\
2013/06/24	&	56468.42	&	18.4	 &$ 	15.15	\pm	0.08	$&$	15.15	\pm	0.11	$&$	14.57	\pm	0.04	$&$	14.19	\pm	0.04	$&$	14.12	\pm	0.07	$&	MEIA	\\
2013/06/24	&	56468.42	&	18.4	 &$ 	15.32	\pm	0.04	$&$	15.18	\pm	0.23	$&$	14.67	\pm	0.08	$&$	14.21	\pm	0.04	$&$	14.18	\pm	0.08	$&	TNG	\\
2013/06/25	&	56469.42	&	19.4	 &$ 	15.14	\pm	0.07	$&$	15.13	\pm	0.13	$&$	14.57	\pm	0.04	$&$	14.17	\pm	0.04	$&$	14.06	\pm	0.06	$&	MEIA	\\
2013/06/26	&	56469.56	&	19.6	 &$ 	15.39	\pm	0.06	$&$	15.22	\pm	0.05	$&$	14.63	\pm	0.05	$&$	  	$&$	  	$&	UVOT	\\
2013/06/26	&	56470.42	&	20.4	 &$ 	15.18	\pm	0.10	$&$	15.20	\pm	0.11	$&$	14.55	\pm	0.03	$&$	14.16	\pm	0.08	$&$	14.02	\pm	0.06	$&	MEIA	\\
2013/06/28	&	56472.44	&	22.4	 &$ 	15.39	\pm	0.05	$&$	15.30	\pm	0.03	$&$	14.55	\pm	0.02	$&$	14.22	\pm	0.12	$&$	14.03	\pm	0.04	$&	AFOSC	\\
2013/06/29	&	56473.42	&	23.4	 &$ 	15.40	\pm	0.07	$&$	15.47	\pm	0.13	$&$	14.62	\pm	0.03	$&$	14.18	\pm	0.04	$&$	14.08	\pm	0.06	$&	MEIA	\\
2013/06/30	&	56474.29	&	24.3	 &$ 	15.78	\pm	0.06	$&$	15.60	\pm	0.05	$&$	14.76	\pm	0.05	$&$	  	$&$	  	$&	UVOT	\\
2013/06/30	&	56474.42	&	24.4	 &$ 	15.69	\pm	0.16	$&$	15.63	\pm	0.14	$&$	14.68	\pm	0.03	$&$	14.22	\pm	0.05	$&$	14.08	\pm	0.07	$&	MEIA	\\
2013/07/02	&	56475.91	&	25.9	 &$ 	16.07	\pm	0.06	$&$	15.87	\pm	0.05	$&$	14.94	\pm	0.05	$&$	  	$&$	  	$&	UVOT	\\
2013/07/03	&	56477.42	&	27.4	 &$ 	16.19	\pm	0.09	$&$	15.98	\pm	0.14	$&$	15.02	\pm	0.04	$&$	14.43	\pm	0.05	$&$	14.26	\pm	0.07	$&	MEIA	\\
2013/07/03	&	56478.13	&	28.1	 &$ 	16.39	\pm	0.08	$&$	16.20	\pm	0.06	$&$	15.11	\pm	0.06	$&$	  	$&$	  	$&	UVOT	\\
2013/07/04	&	56478.42	&	28.4	 &$ 	16.27	\pm	0.10	$&$	16.16	\pm	0.13	$&$	15.15	\pm	0.05	$&$	14.55	\pm	0.07	$&$	14.43	\pm	0.07	$&	MEIA	\\
2013/07/05	&	56479.42	&	29.4	 &$ 	16.48	\pm	0.11	$&$	16.34	\pm	0.13	$&$	15.23	\pm	0.04	$&$	14.55	\pm	0.05	$&$	14.52	\pm	0.04	$&	MEIA	\\
2013/07/05	&	56479.76	&	29.8	 &$ 	16.78	\pm	0.09	$&$	16.32	\pm	0.06	$&$	15.35	\pm	0.06	$&$	  	$&$	  	$&	UVOT	\\
2013/07/06	&	56480.42	&	30.4	 &$ 	16.49	\pm	0.08	$&$	16.48	\pm	0.15	$&$	15.34	\pm	0.06	$&$	14.65	\pm	0.02	$&$	14.50	\pm	0.06	$&	MEIA	\\
2013/07/07	&	56480.58	&	30.6	 &$ 	  	$&$	16.40	\pm	0.16	$&$	15.43	\pm	0.06	$&$	14.60	\pm	0.09	$&$	  	$&	N.Schramm	\\
2013/07/07	&	56481.42	&	31.4	 &$ 	16.66	\pm	0.10	$&$	16.62	\pm	0.14	$&$	15.43	\pm	0.10	$&$	14.73	\pm	0.20	$&$	14.51	\pm	0.04	$&	MEIA	\\
2013/07/08	&	56481.76	&	31.8	 &$ 	16.82	\pm	0.09	$&$	16.63	\pm	0.06	$&$	15.52	\pm	0.06	$&$	  	$&$	  	$&	UVOT	\\
2013/07/09	&	56482.58	&	32.6	 &$ 	  	$&$	16.73	\pm	0.11	$&$	15.52	\pm	0.20	$&$	14.96	\pm	0.25	$&$	  	$&	N.Schramm	\\
2013/07/10	&	56484.25	&	34.2	 &$ 	17.06	\pm	0.10	$&$	16.93	\pm	0.07	$&$	15.84	\pm	0.09	$&$	  	$&$	  	$&	UVOT	\\
2013/07/12	&	56486.33	&	36.3	 &$ 	17.20	\pm	0.14	$&$	17.00	\pm	0.09	$&$	15.73	\pm	0.08	$&$	  	$&$	  	$&	UVOT	\\
2013/07/14	&	56487.64	&	37.6	 &$ 	17.24	\pm	0.12	$&$	17.05	\pm	0.08	$&$	15.79	\pm	0.07	$&$	  	$&$	  	$&	UVOT	\\
2013/07/15	&	56488.58	&	38.6	 &$ 	  	$&$	17.05	\pm	0.03	$&$	15.80	\pm	0.22	$&$	15.26	\pm	0.22	$&$	  	$&	N.Schramm	\\
2013/07/16	&	56489.75	&	39.8	 &$ 	17.33	\pm	0.11	$&$	17.18	\pm	0.10	$&$	16.18	\pm	0.11	$&$	  	$&$	  	$&	UVOT	\\
2013/07/18	&	56492.38	&	42.4	 &$ 	  	$&$	17.19	\pm	0.07	$&$	15.99	\pm	0.03	$&$	15.30	\pm	0.20	$&$	14.93	\pm	0.05	$&	AFOSC	\\
2013/07/20	&	56493.93	&	43.9	 &$ 	17.77	\pm	0.17	$&$	17.33	\pm	0.02	$&$	16.03	\pm	0.09	$&$	15.25	\pm	0.01	$&$	15.18	\pm	0.01	$&	IO:O	\\
2013/07/23	&	56496.68	&	46.7	 &$ 	17.62	\pm	0.11	$&$	17.26	\pm	0.07	$&$	16.19	\pm	0.06	$&$	  	$&$	  	$&	UVOT	\\
2013/07/22	&	56496.42	&	46.4	 &$ 	  	$&$	17.39	\pm	0.19	$&$	16.13	\pm	0.07	$&$	15.37	\pm	0.07	$&$	15.08	\pm	0.08	$&	MEIA	\\
2013/07/22	&	56495.94	&	45.9	 &$ 	17.89	\pm	0.20	$&$	17.34	\pm	0.04	$&$	16.05	\pm	0.04	$&$	15.34	\pm	0.01	$&$	15.16	\pm	0.02	$&	IO:O	\\
2013/07/24	&	56498.40	&	48.4	 &$ 	17.57	\pm	0.10	$&$	17.41	\pm	0.17	$&$	16.13	\pm	0.11	$&$	  	$&$	15.11	\pm	0.08	$&	MEIA	\\
2013/07/27	&	56501.03	&	51.0	 &$ 	17.75	\pm	0.21	$&$	17.50	\pm	0.13	$&$	16.23	\pm	0.11	$&$	  	$&$	  	$&	UVOT	\\
 2013/07/29	&	56503.40	&	53.4	 &$ 	  	$&$	17.46	\pm	0.16	$&$	16.27	\pm	0.04	$&$	15.49	\pm	0.06	$&$	15.17	\pm	0.07	$&	MEIA	\\

\hline
\end{tabular} \\

\label{UBVRI_phot}
\end{table*}

\begin{table*}
\centering
\contcaption{}
\begin{tabular}{ccccccccc} \hline 
    Date & JD & Phase$^{a}$    &    \textit{U}  &  \textit{B} & \textit{V}                & \textit{R} & \textit{I} & Instrument $^{b}$\\ 
         & (+2400000.00)   & (d) & (mag)& (mag)&(mag)&(mag)&(mag)& \\
  \hline

2013/07/30	&	56504.40	&	54.4	 &$ 	  	$&$	17.40	\pm	0.16	$&$	  	$&$	15.56	\pm	0.01	$&$	15.20	\pm	0.05	$&	MEIA	\\
2013/07/31	&	56504.73	&	54.7	 &$ 	17.77	\pm	0.11	$&$	17.50	\pm	0.07	$&$	16.38	\pm	0.07	$&$	  	$&$	  	$&	UVOT	\\
2013/08/06	&	56511.01	&	61.0	 &$ 	17.78	\pm	0.11	$&$	17.54	\pm	0.07	$&$	16.52	\pm	0.07	$&$	  	$&$	  	$&	UVOT	\\
2013/11/20	&	56617.21	&	167.2	 &$ 	  	$&$	18.73	\pm	0.07	$&$	18.28	\pm	0.06	$&$	17.88	\pm	0.04	$&$	17.01	\pm	0.03	$&	IO:O	\\
2013/12/05	&	56632.72	&	182.7	 &$ 	  	$&$	  	$&$	18.42	\pm	0.07	$&$	17.92	\pm	0.05	$&$	17.12	\pm	0.09	$&	AFOSC	\\
2013/12/12	&	56638.68	&	188.7	 &$ 	  	$&$	19.18	\pm	0.04	$&$	18.81	\pm	0.04	$&$	17.97	\pm	0.04	$&$	17.19	\pm	0.12	$&	AFOSC	\\
2013/12/18	&	56645.13	&	195.1	 &$ 	  	$&$	19.10	\pm	0.07	$&$	18.69	\pm	0.04	$&$	18.34	\pm	0.04	$&$	17.38	\pm	0.04	$&	IO:O	\\
2013/12/25	&	56652.28	&	202.3	 &$ 	  	$&$	19.30	\pm	0.07	$&$	18.84	\pm	0.06	$&$	18.51	\pm	0.05	$&$	17.46	\pm	0.03	$&	IO:O	\\
2014/01/08	&	56666.73	&	216.7	 &$ 	  	$&$	19.51	\pm	0.09	$&$	19.23	\pm	0.05	$&$	18.39	\pm	0.03	$&$	17.55	\pm	0.13	$&	AFOSC	\\
2014/01/27	&	56685.21	&	235.2	 &$ 	  	$&$	19.62	\pm	0.08	$&$	19.44	\pm	0.08	$&$	19.10	\pm	0.06	$&$	18.03	\pm	0.05	$&	IO:O	\\
2014/02/11	&	56700.17	&	250.2	 &$ 	  	$&$	20.07	\pm	0.08	$&$	19.91	\pm	0.08	$&$	19.38	\pm	0.08	$&$	18.28	\pm	0.06	$&	IO:O	\\

\hline
\end{tabular} \\
\begin{flushleft}
$^{a}${Phase in days with respect to the adopted explosion date JD = $2456450.0\pm0.9$}\\
$^{b}${F. Ciabattari = 0.5m Newtonian + FLI Proline; S. Donati = 0.3m Schmidt Cassegrain + SBIG ST10;\\ Stan Howerton T18 = 0.25m Telescope + ST10; Koichi Itagaki = Itagaki Astronomical Observatory + KAF-1001E; \\ N. Shcramm = 0.20m Schmidt Newtonian + Orion Star Shoot Pro V2; Stan Howerton T5 = 0.32m Telescope + SBIG STL-6303; MEIA = TJO 0.82m + MEIA; RATCam = LT 2.2m + RATCam;
LRS = TNG 3.58m + LRS; AFOSC = Asiago 1.82m + AFOSC;\\ IO:O = LT 2.2m + IO:O; UVOT = UVOT on board \textit{SWIFT} + MIC\\}
\end{flushleft}

\end{table*}

\begin{table*}
\centering
\caption{Optical Sloan photometry of SN~2013df.}
\begin{tabular}{ccccccc} \hline 
    Date & JD & Phase$^{a}$    &    \textit{u}  &  \textit{r} & \textit{i}   & Instrument$^{b}$ \\ 
         & (+2400000.00) & (d) & (mag)& (mag)&(mag)&\\
  \hline
2013/06/18 &56461.92 & 11.9   &   $ 15.94  \pm   0.02  $&     $14.63  \pm     0.02$ &     $14.78  \pm       0.02$ &    RATCam\\
2013/06/20 &56463.87 &  13.9   &   $                      $&     $14.64  \pm     0.04$ &     $14.74  \pm       0.02$ &    RATCam\\
2013/06/22 &56465.92 & 15.9   &   $ 15.99  \pm   0.32  $&     $14.53  \pm     0.05$ &     $14.58  \pm       0.02$ &    RATCam \\
2013/07/20 &56493.93 & 43.9   &   $ 18.71  \pm   0.07  $&     $15.54  \pm     0.01$ &     $15.52  \pm       0.01$ &    IO:O\\
2013/07/22 &56495.94 & 45.9   &   $ 18.85  \pm   0.01  $&     $15.59  \pm     0.01$ &     $15.52  \pm       0.01$ &    IO:O \\
2013/11/20 &56617.21 & 167.2   &$                          $&$  18.05     \pm   0.03                 $ &$  17.53        \pm   0.02               $ &    IO:O \\
2013/12/18 &56645.13 & 195.1   &$                          $&$  18.50     \pm   0.03                 $ &$  17.92        \pm   0.03               $ &    IO:O \\
2013/12/25 &56652.28 & 202.3    &$                         $&$  18.65     \pm   0.03                 $ &$  18.02        \pm   0.02               $ &    IO:O \\
2014/01/27 &56685.21 & 235.2    &$                         $&$  19.22     \pm   0.04                 $ &$  18.59        \pm   0.04               $ &    IO:O \\
2014/02/11 &56700.17 & 250.2    &$                         $&$  19.58     \pm   0.05                 $ &$  18.85        \pm   0.04               $ &    IO:O \\

\hline
\end{tabular} \\
\begin{flushleft}
$^{a}${Phase in days with respect to the adopted explosion date JD = $2456450.0\pm0.9$}\\
$^{b}$ RATCam = LT 2.2m + RATCam;  IO:O = LT 2.2m + IO:O
\end{flushleft}

\label{uri_phot}
\end{table*}

\begin{figure}
 \includegraphics[width=0.5\textwidth]{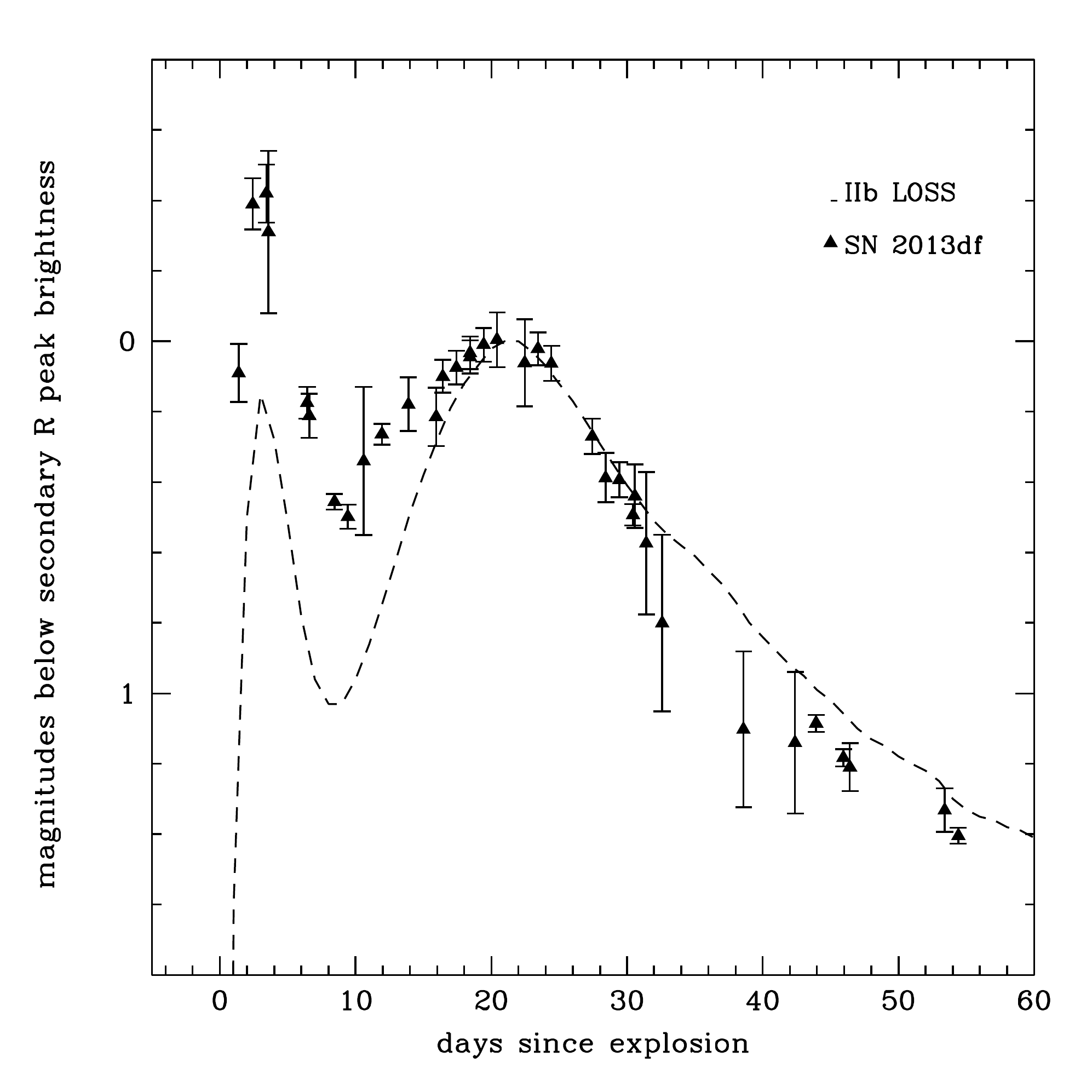}
 \caption{Comparison of the \textit{R} band magnitudes since explosion of SN~2013df and the template for type IIb SNe derived in \citealt{Li11}. }
 \label{fig:templateLOSS}
\end{figure}

To estimate the explosion date of SN~2013df we rely on the similarity of its \textit{R} band light curve with that of SN~1993J. In Figure \ref{fig:templateLOSS} we compared the template for type IIb SN LCs derived from the \textit{R} band data of SN~1993J in \citet{Li11},
and the early time data for SN~2013df. As can be seen in the figure, we perceive the rise to the first maximum in the LC of SN~2013df. Assuming that the rise to the first peak in the \textit{R} band is the same in SN~1993J and SN~2013df, we estimate that the explosion of SN~2013df took place on $\rm JD=2456450.0 \pm 0.9$ (2013 June 6.50 UT). This date is consistent with the discovery date on 2013 June 7.87 UT ($\rm JD = 2456451.37$) and the last non-detection of the SN (2013 May 25 down to a magnitude of 18.5, \citealt{VD13c}). We  also ran comparisons between our sequence of spectra and those in  SNID \citep{snid} and Gelato \citep{gelato} and our estimated explosion date is consistent with the phases of the best fitting spectra found in those databases.

So, we adopt $\rm JD = 245650.0\pm0.9$ as the explosion date of SN~2013df and use it as reference in the rest of this manuscript.\\

The overall shape of the LCs of SN~2013df is similar to those of SNe 1993J and 2011fu. We have estimated the decline rate at the first peak, and the rise and decline rate at secondary peak of SN~2013df through linear interpolation of the observed magnitudes. The results are presented in Table~\ref{decay_rise}. The decline rates after the first peak are in general smaller than those of SN~1993J, and greater than those of SN~2011fu in all bands except \textit{U} (see table 5 of \citealt{11fu}). The 
rise rate to the 
secondary maximum is slower than in SNe 1993J and 2011fu in all bands. However the decline rates after the secondary maximum are faster than in SNe 1993J and 2011fu. The decline rates measured from 
the late time data are steeper than the rate expected for the decay of $^{56}$Co $\rightarrow$
$^{56}$Fe, which is $0.98$ mag (100 d)$^{-1}$. This is a common characteristic to other type IIb SNe and is attributed to the progressive decrease of the $\gamma$ ray trapping in the expanding low mass ejecta.
The minimum magnitudes after first decline and those at secondary maximum of the \textit{BVRI} light curves of SN~2013df were estimated by fitting low order polynomials as were the times at which these minima ($t_{\rm min}$) and maxima ($t_{\rm max}$) took place  (see Table~\ref{min_max}). We were unable to fit the \textit{U} band light curve given its flatness after first decline. With our adopted explosion date of JD = $2456450.0\pm0.9$ for SN~2013df, the minimum occurs at a later time than in SN~1993J and earlier than in SN~2011fu, see table 4 of \citet{11fu}. The difference in magnitude between the minimum and the secondary maximum of the light curves is smaller for  SN~2013df than for SN~1993J and 2011fu. The LCs of all three SNe reach the minimum after the first peak earlier in the redder passbands with the exception of the \textit{I} band. As mentioned before, in the \textit{R} band light curve of SN~2013df, we perceive the rise to the 
first maximum thanks to the early amateur data points.  We have also fitted this first maximum with low order polynomials and estimated that it occurred on $\rm JD_{R1stmax} = 2456453.5\pm0.1$ at $m_{\rm R1stmax} = 13.78\pm0.04$ mag. For SN~1993J, the rise to the first peak in \textit{V} and \textit{R} bands was probed by early observations \citep{Richmond94,Barbon95}, while for SN~2011fu, observations started after the first peak.

  \cite{ecIIb} proposed that type IIb SN LCs presenting an early luminous phase comparable to that of SN~1993J arise from the explosion of extended stars, whereas those lacking an early luminous peak arise from the collapse of more compact stars. For SN~2011dh, which presented an early peak in the \textit{g} band, these progenitor scenarios were explored by modelling the data \citep{
Melina12}. It was shown that a compact progenitor model does not reproduce 
the  early 
time luminous \textit{g} 
observations of the SN whereas an extended model does. For SN~2013df's progenitor, \cite{VD13c} estimated a radius of $545 \pm 65$ R$_{\odot}$ based on their estimates of $T_{\rm eff}$ and $L_{\rm bol}$. In section \ref{subsection:constraints} we set further constraints on the progenitor radius of SN~2013df based on our early photometric data.    \\

\begin{figure}
 \includegraphics[width=0.5\textwidth]{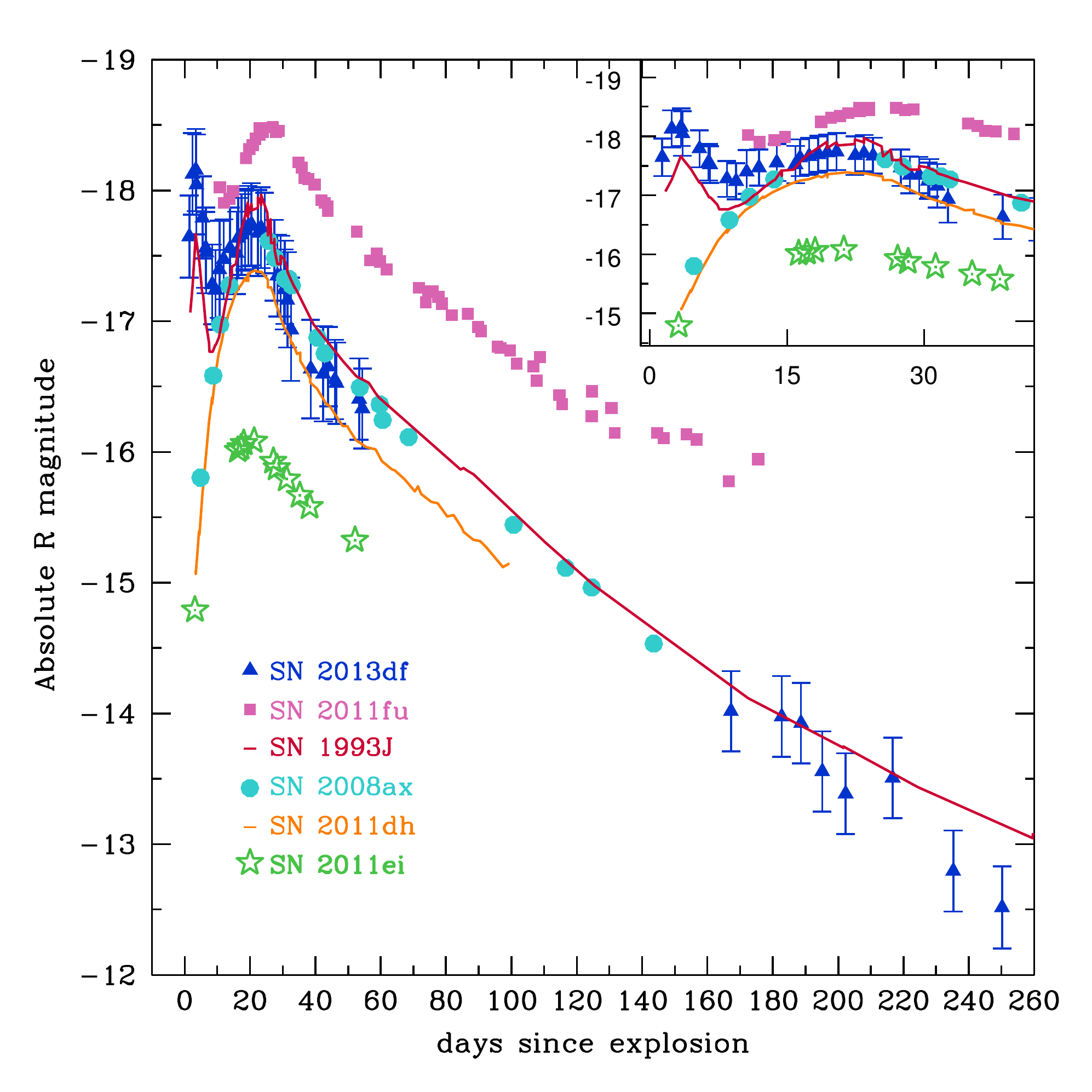}
 \caption{ \textit{R} absolute magnitudes for several type IIb SNe: 2011fu, 1993J, 2008ax, 2013df, 2011dh, 2011ei. For clarity, we have zoomed in on the light curves up to phase $\sim$ 40 d in the upper right corner of the figure (colour figure in the online version). The magnitudes of SN~2013df have been corrected for the assumed reddening $E(B-V)_{{\rm Total}}=0.098\pm0.016$ mag and distance modulus $\mu = 31.65 \pm 0.30$ mag. Apparent magnitudes, extinction, distance and explosion dates to derive the curves for the comparison SNe were taken from the literature (see also Table~\ref{abs-Vmag}).} 
 \label{fig:VabsLCs}
\end{figure}

A comparison of the absolute \textit{R}  LC  of SN~2013df to those of SNe IIb 1993J, 2008ax, 2011dh, 2011ei and 2011fu can be seen in Figure  \ref{fig:VabsLCs}. SN~2013df is dimmer than SN~2011fu and 1993J at the secondary peak, has similar secondary peak brightness as SN~2008ax, and is brighter than SN~2011dh and the low-luminosity type IIb SN~2011ei  (see Table~\ref{abs-Vmag} for a comparison of the \textit{R} absolute magnitudes for different type IIb SNe at maximum).

\begin{table*}
\centering
\caption{Decline and rise rates in the \textit{UBVRI} LCs of SN~2013df.}
\begin{tabular}{ccccc} \hline 
    Band    &    Decline from 1st peak $^{a}$   & Rise to 2nd peak $^{b}$  & Decline from 2nd peak $^{c}$ & Decline tail $^{d}$ \\ 
            &    (mag d$^{-1}$)     &   (mag d$^{-1}$)    & (mag d$^{-1}$)&[mag (100d)$^{-1}$] \\
  \hline
\textit{U}  &$0.15 \pm 0.02$ &       & $0.15 \pm 0.01$ &$2.82\pm 0.68$ \\ 
\textit{B}  &$0.15 \pm 0.02$ &  $-0.03 \pm 0.02$     &$0.13\pm 0.01$  &$1.26\pm 0.02$ \\ 
\textit{V}  &$0.13 \pm 0.01$ &  $-0.06 \pm 0.01$      &  $0.07 \pm 0.01$& $1.81 \pm 0.03$     \\ 
\textit{R}  &$0.16 \pm 0.01$ &   $-0.04 \pm 0.01$     & $0.07 \pm 0.01$ & $1.94 \pm 0.04$ \\ 
\textit{I}  &$0.14 \pm 0.01$ &  $-0.05 \pm 0.01$     & $0.06 \pm 0.01$ &$1.46 \pm 0.05 $ \\

\hline
\end{tabular}
\begin{flushleft}
 $^{a}$ Considering the interval from  $\sim$ 4 to 10 days after explosion \\
 $^{b}$ Considering the interval from $\sim$ 10 to 20 days after explosion \\
 $^{c}$ Considering the interval from $\sim$ 20 to 35 days after explosion \\
 $^{d}$ Considering the interval from $\sim$ 40 days after explosion \\
\end{flushleft}

\label{decay_rise}
\end{table*}

\begin{table*}
\centering
\caption{Minimum and secondary maximum \textit{BVRI} magnitudes of SN~2013df and the corresponding times at which they occurred.}
\begin{tabular}{cccccc} \hline 
    Band    &    $t_{\rm min}^{a}$   &  Apparent min. magnitude   & $t_{\rm max}^{a} $                & Apparent max. magnitude   & Absolute max. magnitudes  \\ 
            &    (d)                  &    (mag)                             &   (days)                              &   (mag)                             & (mag) \\
  \hline

\textit{B}  &$11.27\pm 1.08$ & $15.39\pm0.02$      & $18.46\pm0.91$ & $15.19\pm0.01$ &  $-16.86\pm 0.31$\\ 
\textit{V}  &$10.90\pm  0.90$&  $14.99\pm0.03$      & $20.15\pm1.06$ & $14.57\pm0.02$   & $-17.38\pm 0.30$      \\ 
\textit{R}  &$8.97\pm  0.90$&  $14.69\pm0.30$      & $20.81\pm1.20$ & $14.18\pm0.01$ & $-17.71\pm 0.31$ \\ 
\textit{I}  &$11.02\pm 1.20$ & $14.50\pm0.03$      & $21.60\pm1.25$ & $14.03\pm0.01$  & $-17.80\pm 0.31$  \\

\hline
\end{tabular} 
\begin{flushleft}
  $^{a}$  $t_{\rm min}$ and $t_{\rm max}$ are calculated with respect to our adopted explosion date $\rm JD = 245650.0\pm0.9$.
\end{flushleft}

\label{min_max}
\end{table*}

\begin{table*}
 \setlength{\tabcolsep}{5pt}
 \centering
 \caption{Properties of other type IIb SNe used for comparison along this work.}
 \begin{tabular}{ccccccccl} \hline
  SN & $M_{ R\rm max}^{a}$  & $\mu$& Redshift$^{b}$& \textit{E(B-V)} & $E_{\rm kin}$ & $^{56}$Ni mass & $M_{\rm ej}$ &  Reference \\
  & (mag)& (mag) & &(mag) & ($10^{51}$erg) & (M$_{\odot}$) & (M$_{\odot}$)& \\
   \hline
  1993J & $-17.88 \pm 0.38$ & $27.80 \pm 0.03$ &$-0.00011\pm0.00001$ &$0.19\pm 0.09$& $0.7$ & $0.10$ & $1.3$ &\cite{Richardson06} \\
  1993J & $-17.88 \pm 0.38$ & $27.80\pm 0.03$ &$-0.00011\pm0.00001$ &$0.19\pm 0.09$& $1-1.4$ & $0.10-0.14$ & $1.9-3.5$ &\cite{Young95}\\

  2008ax & $-17.69\pm0.39$ &$29.92\pm0.29$ &$0.00189 \pm 0.00001$ &$0.40\pm0.10$ & $1-6$ & $0.07-0.15$ & $2-5$& \cite{tau08ax}\\

  2011dh & $-17.38^{+0.34}_{-0.29}$ &$29.46^{+0.29}_{-0.27}$ &$0.00200$ &$0.07^{+0.07}_{-0.04}$ & $0.6-1.0$ & $0.05-0.10$ & $1.8-2.5$ & \cite{Ergon11dh100D}\\
  2011ei & $-16.08 \pm 0.44$ &$32.27\pm0.43$ & $0.00931\pm0.00001$&$ 0.24 \pm 0.02$ & $2.5$ & $0.03$ & $1.6$ & \cite{Mili12} \\
  2011fu & $-18.51\pm 0.34$&$34.46\pm0.15$ &$0.01849\pm0.00004$ & $0.22\pm0.11$ & $0.25-2.4$ & $0.21$ & $1.1$ & \cite{11fu}\\
  2013df & $-17.71\pm0.31$ &$31.65\pm0.30
  $ &$0.00239 \pm 0.00002$ &$0.10\pm0.02$ & $0.4-1.2$ & $0.10-0.13$ & $0.8-1.4$ & This work\\
   \hline
 \end{tabular}\\
 \begin{flushleft}
$^{a}${For SNe 1993J, 2011ei and 2011fu we calculated the values of $M_{R\rm max}$ with the apparent magnitudes, distance moduli and extinctions given in the references described in the table.}\\
$^{b}${Host galaxy redshift taken from NED.}\\
\label{abs-Vmag}
 \end{flushleft}

\end{table*}

\subsection{Colour Curves}
\label{subsec:colours}

\begin{figure}
 \includegraphics[width=0.5\textwidth]{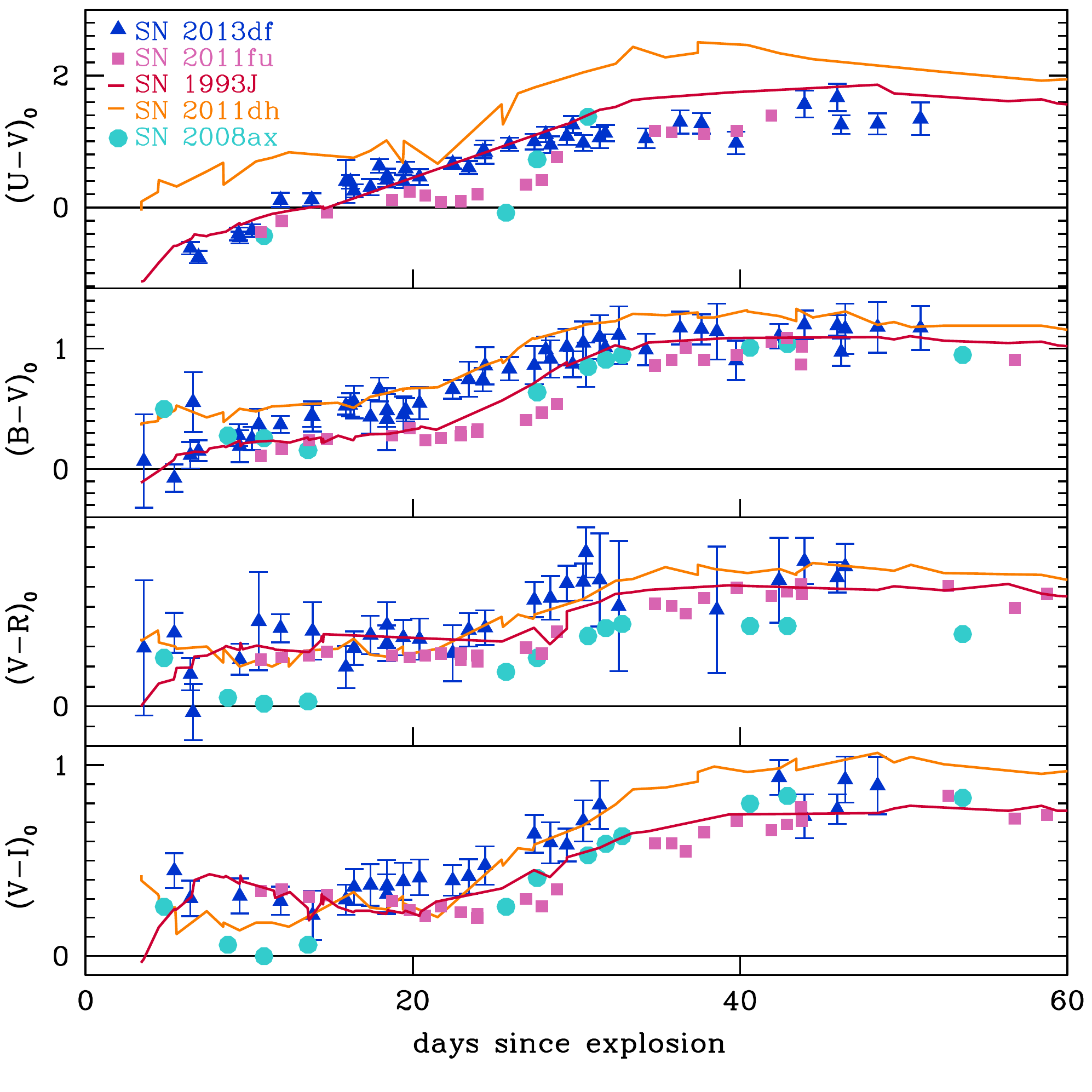}
 \caption{Comparison of the $(U-V)_{0}$, $(B-V)_{0}$, $(V-R)_{0}$, $(V-I)_{0}$ colours of type IIb SNe 1993J, 2008ax, 2011dh and 2013df (coloured in the online version). The colour of SN~2013df has been corrected for the assumed reddening $E(B-V)_{{\rm Total}}=0.098\pm0.016$ mag, while the adopted reddening for the comparison supernovae are given in Table~\ref{abs-Vmag}.}
 \label{fig:colours}
\end{figure}

In Figure \ref{fig:colours} we have plotted the intrinsic colour evolution of SN~2013df together with those of type IIb SNe 1993J, 2008ax, 2011dh, and 2011fu (using the reddening values given in Table~\ref{abs-Vmag} for the comparison SNe). In the first 10 to 15 days after explosion, SN~2013df's $(V-I)_{0}$ colour shows a blue-ward trend which is a common behaviour to all comparison SNe except SN~1993J, whose $(V-I)_{0}$ colour becomes redder. Also during the first 10 to 15 days, both $(U-V)_{0}$ and $(B-V)_{0}$ turn red, specifically the $(B-V)_{0}$ behaviour at this time is opposite to that of SN~2008ax.  There is no clear variation of the $(V-R)_{0}$ index in the first 10 to 15 days. Note that during this same phase for SNe 2008ax and 2011dh $(V-R)_{0}$ slightly decreases, but for SN~1993J this index moderately increases. From 15 days up to about 40 days, all colours become increasingly redder. Between about 40 days and the seasonal gap in observations, there is an evolution 
towards the blue 
in $(U-V)_
{
0}$ and $(B-V)_{0}$. For clarity, in Figure \ref{fig:colours} we have  presented the colour curve comparison only up to 60 days of evolution of the SNe. At our latest observational epochs, SN~2013df shows mean values of $\langle B-V\rangle_{0}=0.23\pm0.03$, $\langle V-R\rangle_{0}=0.46\pm0.12$ and $\langle V-I\rangle_{0}=1.3\pm0.07$. At similar phases, SN~2008ax shows  $\langle V-I\rangle_{0}=0.9$ while for SN~1993J $\langle V-I\rangle_{0}=0.5$, so SN~2013df seems to exhibit redder colours at later phases.
\\

The similarity in the light curves of SNe 1993J and 2013df during the first 10 days after explosion (approximately the duration of the first peak and thus, the time that the LCs are adiabatically powered; \citealt{Ws94}) is mostly reflected in the $(U-V)_{0}$ and $(B-V)_{0}$ indices.  As noted above, the $(V-I)_{0}$ evolution is the opposite for these two SNe during this period, however, there is no clear similar or completely distinct behaviour in their $(V-R)_{0}$ index evolution. During the first 10 days, SN~2011dh's \textit{UBVRI} LCs are radioactively powered \citep{Melina12,Ergon11dh100D}, and present redder $(U-V)_{0}$  and $(B-V)_{0}$ indices with a flatter evolution during this time. This could mean that the main contribution of the shock wave powering the first peak in SNe 1993J and 2013df is in the bluer wavelengths.

\subsection{Bolometric Light Curve}
\label{subsec:bol}

 To derive the UV-optical-NIR pseudo-bolometric LC of SN~2013df, we followed two steps. In the first place, we derived the UV-optical pseudo-bolometric LC from the photometry presented in Section \ref{subsection:UVopt}. This was done calculating the fluxes at the effective wavelengths from the extinction corrected apparent magnitudes. When at a certain epoch there was no observation, it was obtained by interpolation of the LC. The spectral energy distribution given by these effective fluxes was integrated following a trapezoidal rule, and the integrated flux was converted into luminosity for our adopted distance. The next step was to estimate a NIR contribution to the LC, since  SN~2013df was not observed in these wavelengths. To do this, we calculate two pseudo-bolometric LCs of  SN~1993J (using the photometric data from \citealt{Richmond94} and \citealt{Matthews2002}). Following the same procedure as for the UV-optical pseudo-bolometric LC of SN~2013df, one calculation was done considering, and the 
other omitting, the NIR 
contribution. The difference in luminosity between SN 1993J's optical and optical-NIR  pseudo-bolometric LCs was then 
added to SN~2013df's UV-optical LC, thus obtaining its  UV-optical-NIR pseudo bolometric LC. If instead of using the NIR contribution estimated for SN~1993J we use the one obtained for SN~2011dh from the data of \citealt{Ergon11dh100D}, the shape of the UV-optical-NIR pseudo bolometric LC of SN~2013df does not change, however,  the LC results a little less brighter around the minimum after first peak and at secondary peak. The only explosion parameter affected by this is then the $^{56}$Ni mass, which has a slightly lower uncertainty: $ 0.1-0.11$ M$_{\odot}$.

 SN~2013df's luminosities at the minimum after the first peak and at the secondary maximum were estimated by fitting low order polynomials to the light curve. This resulted in $L_{{\rm min}}=2.2\times10^{42}$ erg s$^{-1}$ and $ L_{{\rm 2nd\;max}}=2.5\times10^{42}$ erg s$^{-1}$ respectively. We also derived a lower limit to the luminosity at first peak from our \textit{R} band observations $ L_{{\rm 1st\;peak}}\gtrsim2.5\times10^{42}$ erg~s$^{-1}$.

\begin{figure*}
 \includegraphics[width=12.6cm]{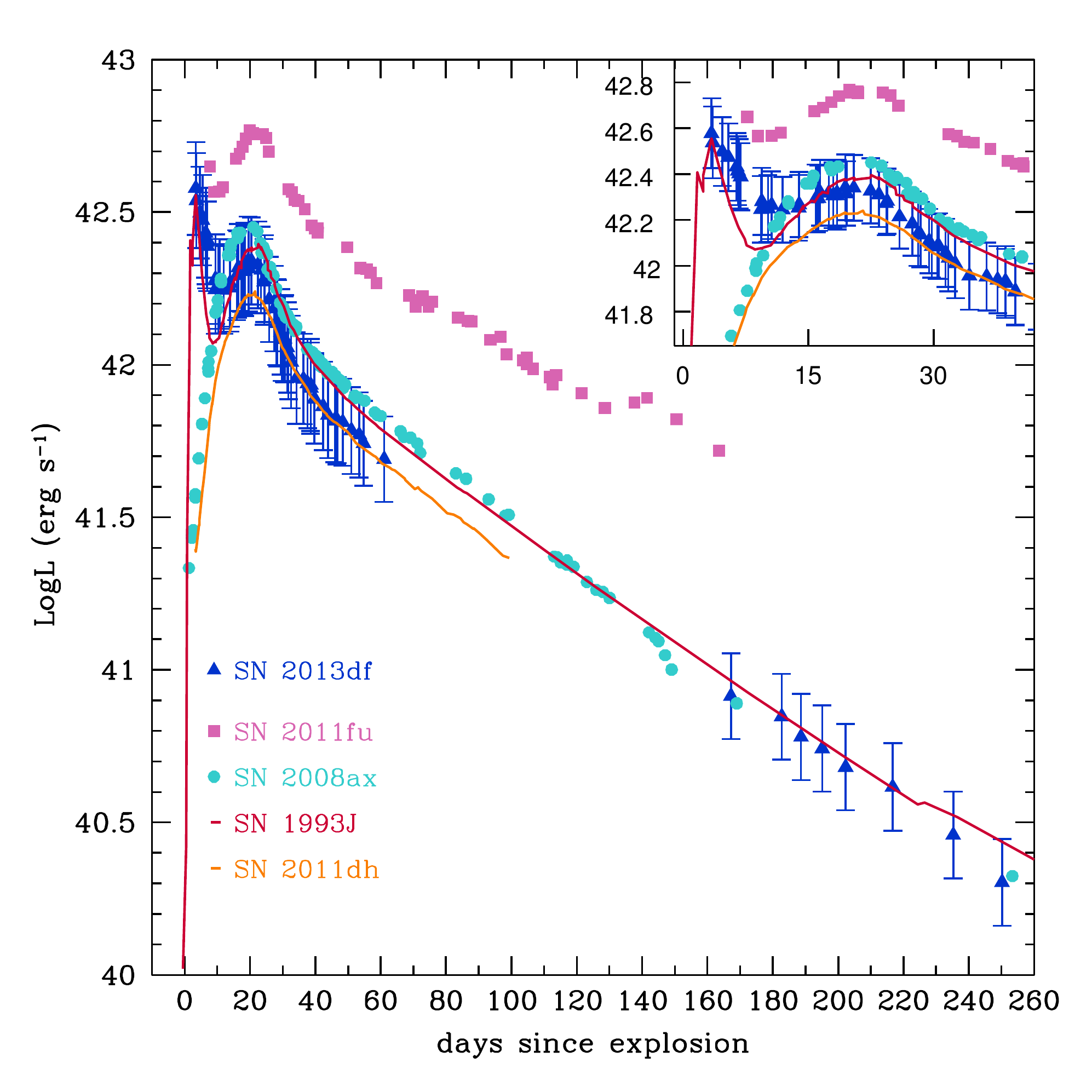}
 \caption{Pseudo-bolometric optical-NIR LC of SN~2013df compared to those of other type IIb SNe (coloured in the online version). For clarity,  we have zoomed in on the light curves up to phase $\sim$ 40 d in the upper right corner of the figure.} 
 \label{fig:bolom_other}
\end{figure*}

\begin{figure}
 \includegraphics[width=0.5\textwidth]{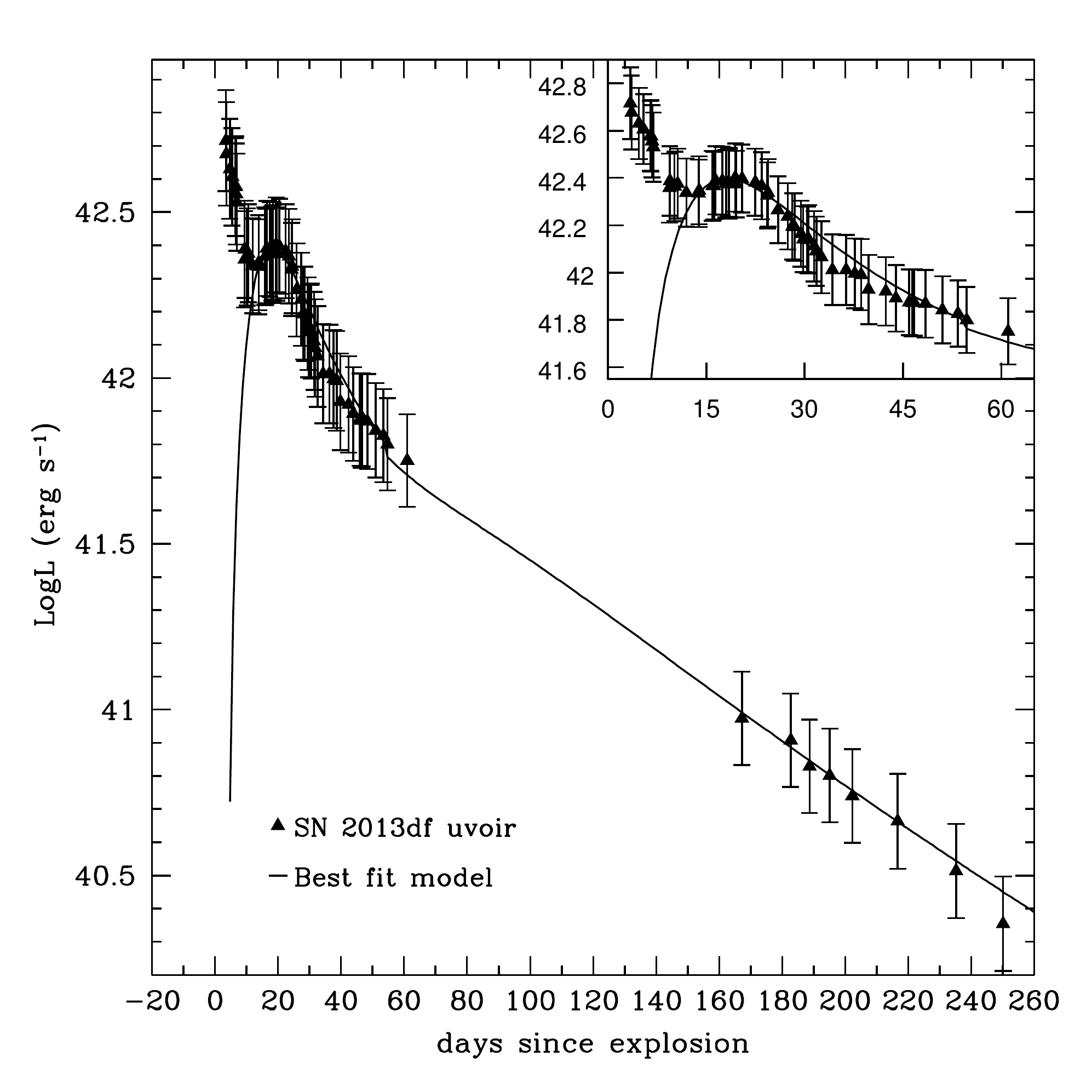}
 \caption{Pseudo-bolometric UV-optical-NIR LC of SN~2013df (calculated assuming the same NIR contribution as SN~1993J)  and  its best fit model computed omitting the first peak. For clarity, we have zoomed in on the light curves up to phase $\sim$ 70 d in the upper right corner of the figure.} 
 \label{fig:bolom}
\end{figure}

In Figure \ref{fig:bolom_other} we present a comparison of the computed optical-NIR  pseudo-bolometric light curves of several type IIb SNe with that of SN~2013df. As can be seen in the figure, SN~2013df has a slower cooling phase after the first peak than SN~1993J, and is the double-peaked SN IIb with the least difference between the minimum after first peak and the secondary maximum. The secondary peak luminosity of SN~2013df lies between those of SN~2011dh and SN~1993J, while the LC width seems smaller following standard light curve interpretation \citep[e.g.,][]{Arnett82}. This suggests that the ejected $^{56}$Ni mass of SN~2013df should lie between those of SNe 2011dh and 1993J while the ejecta mass should be lower than for those two SNe. At late phases, the LC tail follows the trend of SN~1993J 
although the slope is a bit steeper and this again could indicate a lower ejecta mass than for 
SN~1993J.\\
In order to derive the explosion parameters of SN~2013df, we have modelled the bolometric light curve as described in \cite{valenti08}. The SN luminosity evolution is divided into the photospheric phase (t$\leq 30$ d) and the nebular phase (t$\geq 60$ d). The simple analytical model by \cite{Arnett82} is adopted, and in addition we include the energy produced by the $^{56}$Co $\rightarrow$ $^{56}$Fe decay. In the nebular phase, the luminosity is powered by the energy 
deposition of the $\gamma$ rays produced by the  $^{56}$Co decay, the electron-positron annihilation and the kinetic energy of the positrons \citep{Sutherland84,Cappellaro97}. Since the model cannot reproduce shock breakout that dominates the early phases after explosion, we do not attempt to fit the early LC. We limited the photospheric phase to 25 days after the secondary peak, assumed a constant optical opacity $\kappa_{\rm opt}=0.10$ cm$^{2}$g $^{-1}$, and adopted a range of photospheric velocities (derived from the minimum of the Fe\,{\sc ii} spectral lines) spanning $7000$ to $9000$~km s$^{-1}$. The pseudo-bolometric UV-optical-NIR light curve of SN~2013df, as well as 
its best fit model, 
are 
presented in Figure \ref{fig:bolom}, whereas the explosion parameters derived from the model are displayed in Table~\ref{abs-Vmag}. As can be seen in Figure \ref{fig:bolom}, the model overall reproduces the behaviour of the observational data points and the derived $^{56}$Ni and ejecta masses are consistent with the other type IIb SNe, at least if the $^{56}$Ni mass for SN~2011dh is rather below the upper limit derived by \cite{Ergon11dh100D} ($<0.1$ M$_{\odot}$), and for SN~1993J it is in the range $0.1-0.14$ M$_{\odot}$. We measured the width of both the observed and model bolometric LCs, and found that close to peak the widths are very similar but the observed LC is wider by one day. However, as can be seen in Figure \ref{fig:bolom} the model seems broader at around phase 40. Most likely this is related to the fact that it cannot accurately reproduce fast declining LCs, and may indicate that the ejecta mass derived from the model may be slightly overestimated.

\section{Spectroscopic Results}

\subsection{Spectral evolution}
\label{subsection:spec}

\begin{table*}
\centering
\caption{Log of spectroscopic observations of SN~2013df.}
\begin{tabular}{ccccc} \hline 
    Date    &    JD   & Phase $^{a}$  & Set-up$^{b}$& Spectral Range   \\ 
            &  (+2400000)                &   (d)  & & \AA{}                           \\
  \hline
 2013/06/15 &56459.41 &   9.4    &  AFOSC+g4,VPH6& 3500-10000 \\ 
 2013/06/20 &56464.35 &   14.4   &  B\&C+g300& 3350-7850\\ 
 2013/06/28 &56472.41 &   22.4   &  AFOSC+VPH6 & 4500-10000     \\ 
 2013/07/18 &56492.35 &   42.2   &  AFOSC+g4 & 3500-8450\\ 
 2013/12/05 &56632.66 &   182.7  &  AFOSC+g4,VPH6& 3500-10000 \\ 
 2014/02/04 &56693.25&    243.3  &  OSIRIS+R500R& 4800-10000\\

\hline

\end{tabular} \\
\begin{flushleft}
$^{a}${Phase in days with respect to the adopted explosion date JD = $2456450.0\pm0.9$}\\
$^{b}${AFOSC = Asiago 1.82m + AFOSC; B\&C = Asiago 1.22m + B\&C; OSIRIS = Gran Telescopio de Canarias 10.4m + Osiris}

\end{flushleft}

\label{spec_log}
\end{table*}

\begin{figure*}
 \includegraphics[width=15.6cm]{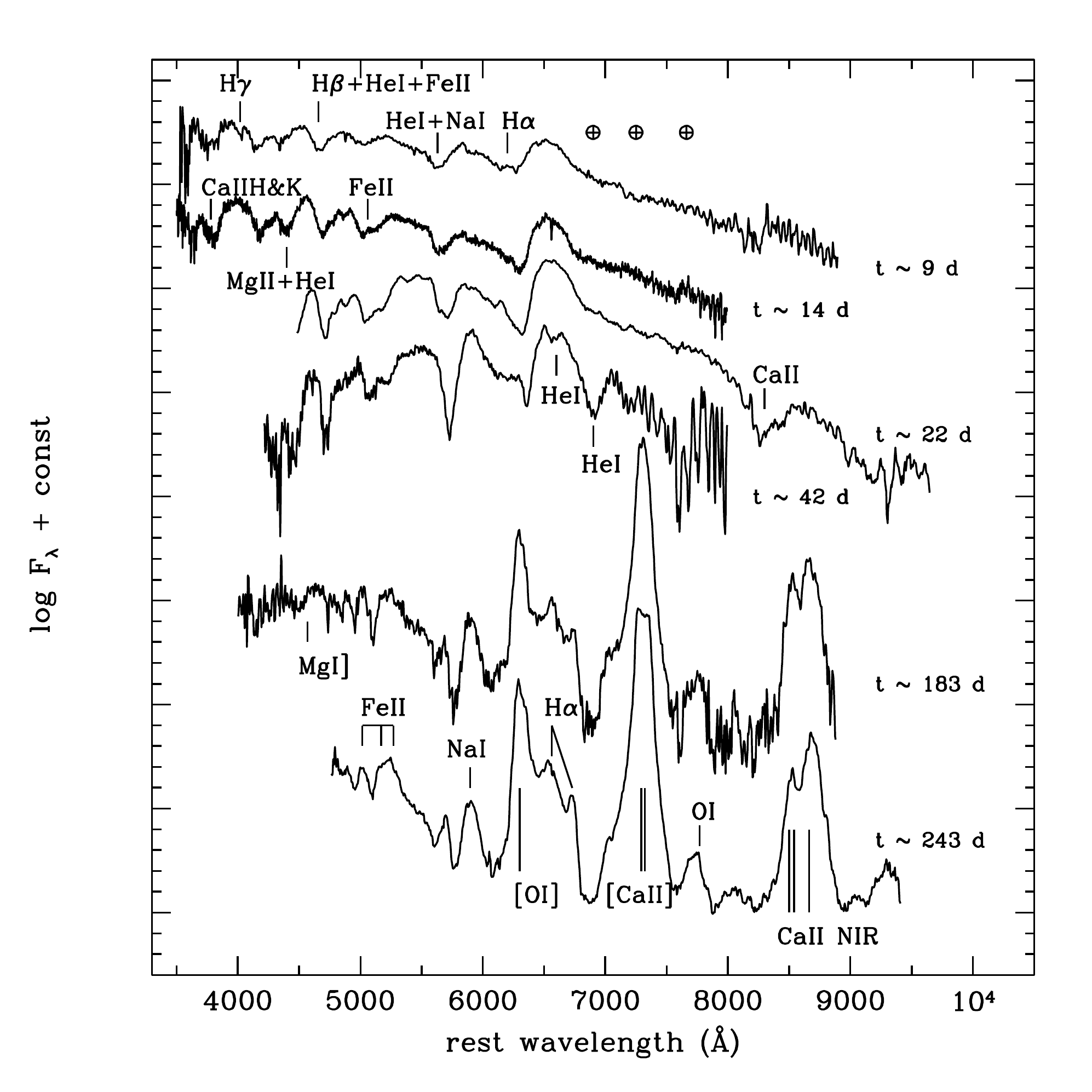}
 \caption{Optical spectral evolution of SN~2013df, where the most relevant features in the spectra are indicated. Telluric features have been  marked with $\oplus$. The spectra have been corrected for the host galaxy redshift. Epochs indicated in the plot are with respect to our assumed explosion date of JD = $2456450.0\pm0.9$. Spectra have been shifted vertically for clarity. }
 \label{fig:specevol}
\end{figure*}

The log of our spectral observations is presented in Table~\ref{spec_log}.
Our spectral sequence of SN~2013df (Figure \ref{fig:specevol}) covers the evolution from 9 up to 243 days post explosion, with a seasonal visibility gap between 42 and 183 days. The most prominent lines are identified in the figure.

In the spectrum at 9 days (that is near the minimum after the first peak in the LCs), we identify a clear P-Cygni H$\alpha$ feature and an absorption component of H$\beta$ at around $4700$~\AA{}, although the latter possibly has contributions of Fe\,{\sc ii} $\lambda$$4924$ \citep{Barbon95,tau08ax,Hach12}. Both the absorption components of H$\alpha$ and H$\beta$ exhibit a flat-bottom profile indicating the presence of a thick expanding H shell \citep{Barbon95}. Other than the flat bottom, and as seen before in type IIb SNe~1993J, 2001ig (\citealt{Silv09}), 2011dh (\citealt{Marion13}) and 2011ei, the H$\alpha$ absorption seems to have two components which may be due to the presence of two peaks in the radial distribution of the H density or to contamination by Si\,{\sc ii}. The spectrum also shows He\,{\sc i} $\lambda$$5876$ absorption probably blended with Na\,{\sc i} $\lambda$$5890-5896$ \citep{tau08ax}. The absorption at approximately $5000$~\AA{} is likely a blend of Fe\,{\sc ii} 
$\lambda$$5018$, He\,{\sc i} 
$\lambda$$5015$ and Si\,{\sc ii} $\lambda$$5016$ \citep{Hach12,Silv09}, and that around $4300$~\AA{} a blend of Mg\,{\sc ii} $\lambda$$4481$ and He\,{\sc i} $\lambda$$4471$ \citep{Hach12,Silv09}. We believe the feature around $4100$~\AA{} could be H$\gamma$,  
and there are also traces of Ca\,{\sc ii} H \& K at $3934$~\AA{}  \& $3968$~\AA{} \citep{tau08ax,Hach12}.

The spectrum taken at 14 days past explosion exhibits the same features as our previous spectrum. However, the features identified as H$\alpha$ and H$\beta$ have increased by a factor in equivalent width of approximately 1.2 with respect to our previous spectrum, while the feature identified as He\,{\sc i} $\lambda$$5876$ has diminished approximately by the same quantity. 
The spectrum taken at 22 days since explosion (near to the secondary maximum light), is characterised by an increase in intensity of the $\lambda$$5876$ He\,{\sc i} line by a factor of about 1.4 while the increase of the H$\alpha$ and H$\beta$ lines is by a factor of approximately 1.1. This spectrum also shows conspicuous absorptions of Ca\,{\sc ii} NIR $\lambda$$\lambda$$\lambda$$8498$, $8542$, $8662$ \citep{tau08ax,Hach12}.

The spectrum at 42 days presents a significant decrease in flux at blue wavelengths compared to earlier data. The H$\alpha$ line diminished by a factor of 4 approximately, while He\,{\sc i} lines increased in intensity with respect to the preceding spectra by a factor of 1.3 approximately. Furthermore, the He\,{\sc i} features at $\lambda$$6678$ and $\lambda$$7065$ respectively, can be identified in this spectrum \citep{tau08ax} while they were absent in the previous spectra. The profiles of the absorption lines of He\,{\sc i} $\lambda$$5876$ and H$\alpha$ seem to be similar to the ones seen in the type IIb SN~2000H at phases 19 and 30 days \citep{Branch02}.

\begin{figure}
 \includegraphics[width=0.5\textwidth]{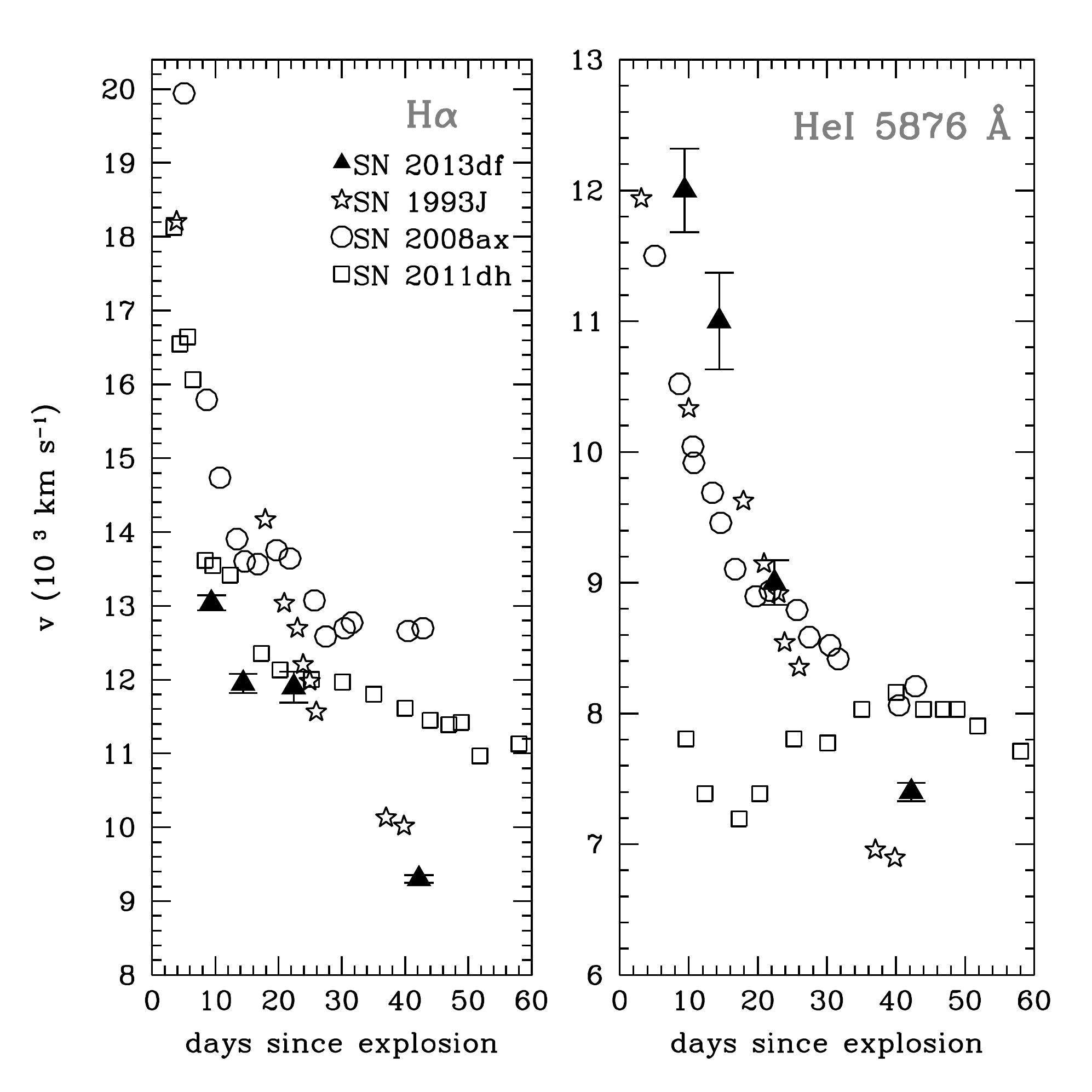}
 \caption{Early evolution of the velocities of H$\alpha$ and He\,{\sc i} $\lambda$$5876$ for IIb SNe 2013df, 1993J and 2008ax and 2011dh.}
 \label{fig:velo}
\end{figure}

The last two spectra of our sequence, taken at phases of 183 and 243 days after explosion, exhibit a relative increase in flux at blue wavelengths as well as strong lines of [O\,{\sc i}] $\lambda$$\lambda$ $6300$, $6364$, [Ca\,{\sc ii}] $\lambda \lambda$$7291$, $7324$, O\,{\sc i} $\lambda$$7774 $, and Ca\,{\sc ii}  NIR. We also identify Na\,{\sc i} around 5890\AA{} possibly contaminated by residual He\,{\sc i} $\lambda$5876, and Fe\,{\sc ii} around $5000$~\AA{} \citep{Matheson2000b,tau08ax,shivvers2013}. We believe the features around $6500$-6600\AA{} and $6700$-6800\AA{} might be associated to H$\alpha$ emission, we discuss this possibility below. The [Ca\,{\sc ii}]  $\lambda \lambda$$7291$, $7324$ feature is noticeably more intense than [O\,{\sc i}] in both of our last spectra.\\
 The work by \cite{WH07} on the nucleosynthesis in massive stars of solar metallicity, shows that oxygen production  strongly increases for progenitors over $\sim$ 16 M$_{\odot}$. In  \citealt{jerkstrand14}, a comparison between the observational and modelled [O\,{\sc i}] $\lambda$$\lambda$$6300$, $6364$  luminosities relative to the $^{56}$Co decay power for SNe 1993J, 2008ax, and 2011dh is shown. All of the SNe exhibit luminosities consistent with progenitors of initial masses ranging 12 to 17 M$_{\odot}$.  We have attempted to do a rough estimate of SN 2013df's progenitor mass by comparing the luminosities derived from [O\,{\sc i}] $\lambda$$\lambda$$6300$, $6364$ in our nebular spectra with the model tracks for type IIb SNe represented in figure 12 of \citealt{jerkstrand14}. In order to calculate such luminosities, first we deblended the emission feature of [O\,{\sc i}] $\lambda$$\lambda$$6300$, $6364$. Then, the integrated flux was obtained adjusting a Gaussian to the feature 
and subtracting the continuum flux level. We converted the resulting flux in luminosity using our adopted distance and the extinction estimated, and then normalized this value relative to the $^{56}$Co decay power (see equation 1 of \citealt{jerkstrand14}). Therefore, we derived $L_{\rm norm}(183)=0.0088^{+0.0020}_{-0.0010} $ and $L_{\rm norm}(243)=0.0075^{+0.0025}_{-0.0013} $ at 183 days and 243 days respectively. These two values are found to be closer to those for models 12C and 13D at similar phases (see figure 12 of \citealt{jerkstrand14}). The main properties of these models can be seen in table 3 of the mentioned manuscript. Still, we stress that $M_{\rm ZAMS}$ for these models are 12$\rm M_{\odot}$ and 
13$\rm M_{\odot}$. These values are consistent with the lower limit for SN 2013df's progenitor mass derived by \cite{VD13c} from the analysis of the HST pre-explosion images of the SN site. SN 2013df has luminosities most similar to those of SNe 2011dh and 2008ax, for whom the values are also consistent with 12-13 $\rm M_{\odot}$ progenitors, while they are quite below the ones for SN 1993J which match best with a higher mass progenitor (M$\sim$15$\rm M_{\odot}$). \\ 
In Figure \ref{fig:velo} we show the evolution of the expansion velocities of H$\alpha$ and He\,{\sc i} $\lambda$$5876$, derived from the positions of the minima of the P-Cygni absorptions, for SNe 2013df, 1993J, 2008ax, and 2011dh. The H$\alpha$ velocity for SN~2013df at 9 days was measured from the position of the red-edge of the flat bottom absorption component as described in \cite{jefbranch90}. The velocities for SNe 1993J and 2008ax were taken from \cite{tau08ax}, while those for SN~2011dh were obtained from \cite{Ergon11dh100D}. The trend of the velocity of He\,{\sc i}  during the first 40 days of evolution of SN~2013df is similar to that of SNe 2008ax and 1993J, but the values are overall higher than for those SNe, and there is a steeper gradient in velocity. As we mentioned above, the line we have identified as He\,{\sc i} $\lambda$$5876$ might be contaminated by Na \,{\sc i}, this fact could have affected the velocities estimated for SN~2013df. SN~2011dh presents quite a 
different evolution of the He\,{\sc i} $\lambda$$5876$ velocities than the rest of the IIb SNe of the comparison. \cite{Ergon11dh100D} pointed out that detailed modelling including non thermal 
excitations might 
provide an explanation for the behaviour of the He lines in SN~2011dh. The trend  of the H$\alpha$ velocity evolution for SN~2013df is similar to that of the comparison SNe but the actual values 
are smaller. We note that if in our first spectrum we measure the minimum of the P-Cygni of the H$\alpha$ line adjusting a Gaussian to the whole profile, we derive a velocity of about 16000 km s$^{-1}$, which is close to the values obtained for the other IIb SNe at similar phase.

We have measured the black-body temperature by fitting the continuum of our spectral sequence. The temperature increases slowly from 7700K at 9 days to 7900K at 14 days, and subsequently decreases to 6900K at 22 days and 4000K at 42 days past explosion. Considering that the errors of these measurements are about $\pm500$K, we could say that the black-body temperature remained constant between 9 and 22 days and decreased after that. Our values are slightly below those reported for SN~2011dh at coeval epochs \citep{Ergon11dh100D}. This means the black-body radius for SN~2013df, interpreted as the thermalization radius, should be greater than for SN~2011dh given that the two SNe have similar luminosity.  

\begin{figure}
 \includegraphics[width=0.5\textwidth]{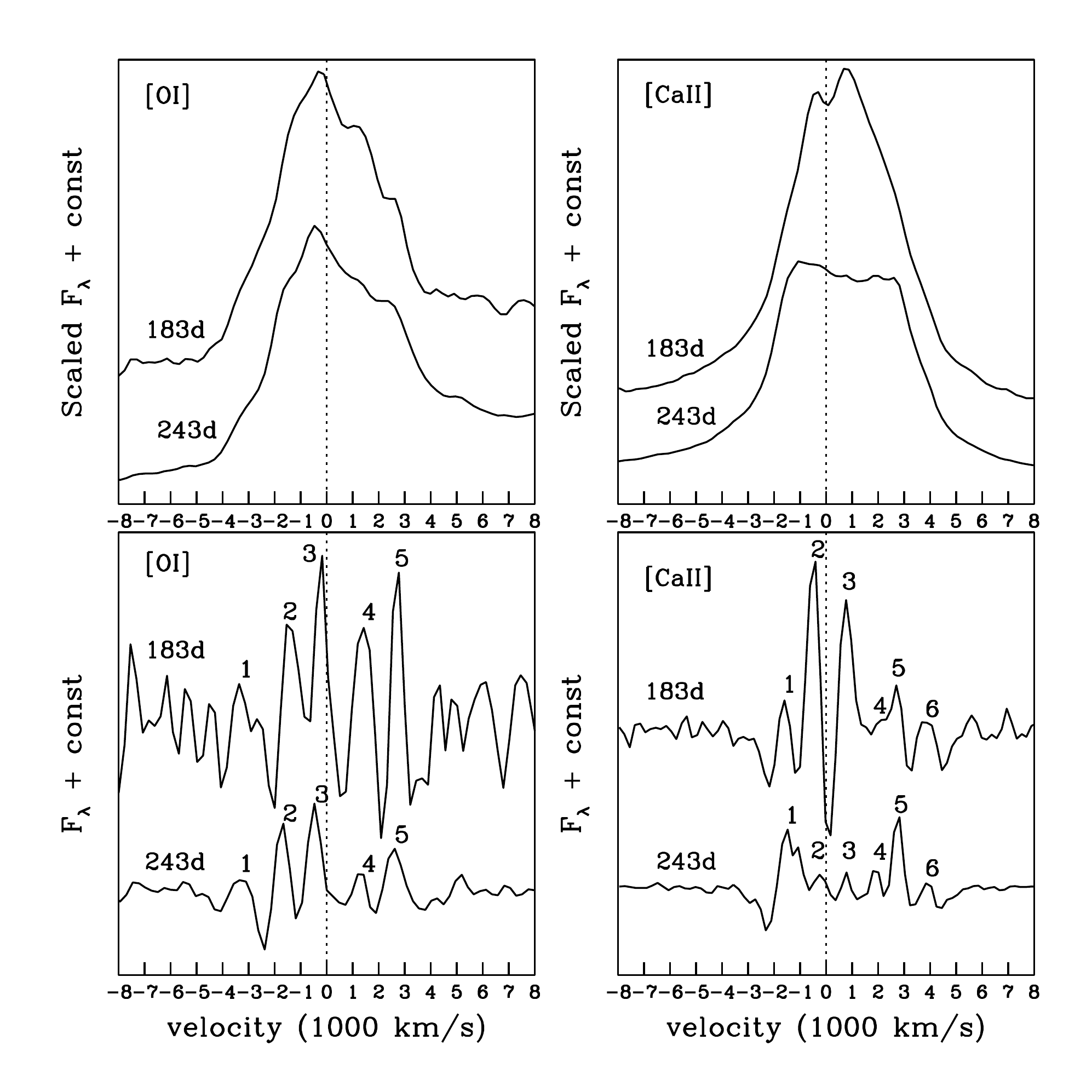}
 \caption{Top panel: Nebular profiles of SN~2013df's [O\,{\sc i}] $\lambda\lambda$$6300$, $6364$  and  [Ca\,{\sc ii}] $\lambda\lambda$$7291$, $7324$ lines at phases 183 and 243 days in velocity space. Bottom panel: Substructures present in the nebular profiles of the [O\,{\sc i}] $\lambda\lambda$$6300$, $6364$  and  [Ca\,{\sc ii}] $\lambda\lambda$$7291$, $7324$ lines at phases 183 and 243 days. The vertical lines at 0 km~s$^{-1}$ mark zero velocity with respect to $6300$ and  $7291$~\AA{}. The numbering in the lower panel corresponds to the different components identified in the profiles.}
 \label{fig:nebularclumps}
\end{figure}

In the top panel of Figure \ref{fig:nebularclumps}, we display the nebular profiles for the [O\,{\sc i}] and the [Ca\,{\sc ii}]  
 ions of SN~2013df in velocity space. Contrary to SN~2008ax and similar to SN~2011dh and 1993J, SN~2013df does not show a symmetrical double-peaked [O\,{\sc i}] profile, but does seem to show some substructure indicative of clumping possibly related to  Rayleigh-Taylor instabilities during the expansion of the SN ejecta \citep[see e.g.][]{Kifonidis2000}. We note that the centroids of [O\,{\sc i}] and [Ca\,{\sc ii}] in our spectra are slightly blue-shifted. This behaviour has been seen in other stripped envelope type Ib/c SNe and could result from residual opacity in the ejecta core \citep{tau09}. We have followed the procedure described by \cite{Matheson2000a} to distinguish the possible different components forming the profiles. In the first place, we smoothed the profiles with boxcars of 5 pixels, and secondly we subtracted the smoothed profiles from the original ones. In the bottom panel of Figure \ref{fig:nebularclumps}, we zoom in on 
the small scale fluctuations we have obtained for [O\,{\sc i}] $\lambda$$6300$, $6364$ and [Ca\,{\sc ii}] $\lambda\lambda$$7291$, $7324$. The dotted vertical lines marking 0 velocity in both 
panels of the plot
refer to $\lambda$$6300$ for [O\,{\sc i}] and $\lambda$$7291$ for [Ca\,{\sc ii}]. We identify 5 components in [O\,{\sc i}] in both the 183 and 243 day spectra, while the [Ca\,{\sc ii}] line seems to consist of 6 features. For [O\,{\sc i}], components 1 through 5 lie at approximately $-3400$, $-1700$, $-500$, $1300$, $2600$ km s$^{-1}$ respectively, and components 1 through 6 of [Ca\,{\sc ii}] are at approximately $-1500$, $-200$, $800$, $1900$, $2750$, $3900$ km s$^{-1}$. Components 3, 4, and 5 in the [O\,{\sc i}] $\lambda$$6364$ velocity space are shifted approximately by the same velocity as components 1, 2, and 3 in the [O\,{\sc i}] $\lambda$$6300$ velocity space, which means that they are probably repetitions. The features that seem to be repeated in the [Ca\,{\sc ii}] profile are 2, 4, and 6 as they are shifted by approximately the same velocity as components 1, 3, 5 if the line is represented in the [Ca\,{\sc ii}] $\lambda$$7324$ velocity space. \\
We performed the same subtraction described above to try to extract the substructure present in the O\,{\sc i} $\lambda$$7774$ line but were only able to distinguish one feature shifted $-500$~km~s$^{-1}$ in our spectrum at phase 243. We were unable to identify any small scale fluctuations for [O\,{\sc i}] $\lambda$$5777$ and for the Mg\,{\sc i}] $\lambda$$4571$ line profiles owing to the low S/N ratio of these features. For SN~1993J, \cite{Matheson2000a} found matches between the small scale fluctuations in the [O\,{\sc i}]  $\lambda \lambda$$6300$, $6364$, [O\,{\sc i}] $\lambda$$5777$ and O\,{\sc i} $\lambda$$7774$  emission lines. However, no match was detected between the features present in  Mg\,{\sc i}] $\lambda$$4571$, [O\,{\sc i}]  $\lambda \lambda$$6300$, $6364$, the various O\,{\sc i} lines and [Ca\,{\sc ii}] $\lambda\lambda$$7291$, $7324$ lines. This was interpreted as the emission of different species forming at different locations. Unfortunately, we cannot confirm whether there is a correlation 
between 
the 
different O\,{\sc i} lines for SN~2013df, but we have found that 
the substructures in [O\,{\sc i}] $\lambda \lambda$$6300$, $6364$ and [Ca\,{\sc ii}] $\lambda\lambda$$7291$, $7324$ are not correlated and probably originate in different clumps.\\

\subsection{Late time emissions between 6500 and 6800~\AA{} }
In our nebular spectra of SN~2013df there are two interesting emission lines. We speculate that these lines could possibly be different components of H$\alpha$ emerging after interaction with circumstellar material (CSM) material. This was the interpretation for the 6600~\AA{} band observed in the 367 day spectrum of SN~1993J \citep{patat95}, as well as in later time spectra \citep{Matheson2000b}. 

One possibility to explain the two distinct components of the nebular H$\alpha$ emission in SN~2013df is that the ejecta-CSM interaction is asymmetrical. For example, triple peaked H$\alpha$ and other Balmer line emission was observed in the spectra of the type IIn SN~2010jp starting at 75 days \citep{
smith2012}. In that case it was claimed that the two Doppler-shifted components of the Balmer lines were caused by circumstellar interaction of a tilted collimated bipolar jet produced during the explosion. Note that in the spectra of SN~2013df, if we consider the feature between $6500$ and 6600~\AA{} to be the central component of the H$\alpha$ emission, and the feature between 6700 and 6800~\AA{} to be the redshifted one, there could possibly be a blue-shifted  H$\alpha$ emission as well but it might be submerged in the [O\,{\sc i}] $\lambda\lambda$$6300$, $6364$ line. Unfortunately, we do not clearly detect other Balmer lines in our nebular spectra, and cannot  confirm the multicomponent profile.   \\ In a recent article on late time line formation in IIb SNe, \citealt{jerkstrand14} stress the fact 
that at phases above 
150 days, no H$\alpha$ emission is produced by their models. Instead the features around 6550~\AA{} are likely caused by [N\,{\sc ii}] $\lambda\lambda$$6548$, $6584$. So, another possibility is that most of the 6500~\AA{} emission is coming from [N\,{\sc ii}] and that there is multi-peaked H$\alpha$ from CSM interaction on top. Although in principle we could attribute the $6700$-6800~\AA{} 
feature to residual He\,{\sc i} $6678$~\AA{}, \citealt{jerkstrand14} stress that optical He lines start diminishing in spectra of type IIb SNe after 100 days and are difficult to detect. So, given the similarity with SN~1993J, and the fact that for this SN the nebular emission around 6600\AA{} was well modelled by H$\alpha$ excited by CSM interaction, we believe the 
emissions that are detected at 
6500-6600 and 6700-6800~\AA{} in SN~2013df's late time spectra could also be caused by multiple component H$\alpha$, and that the 6500-6600\AA{} feature is possibly on top of [N\,{\sc ii}] $\lambda\lambda$$6548$, $6584$.

\subsection{Spectral comparison to other SNe}

\begin{figure}
 \includegraphics[width=0.5\textwidth]{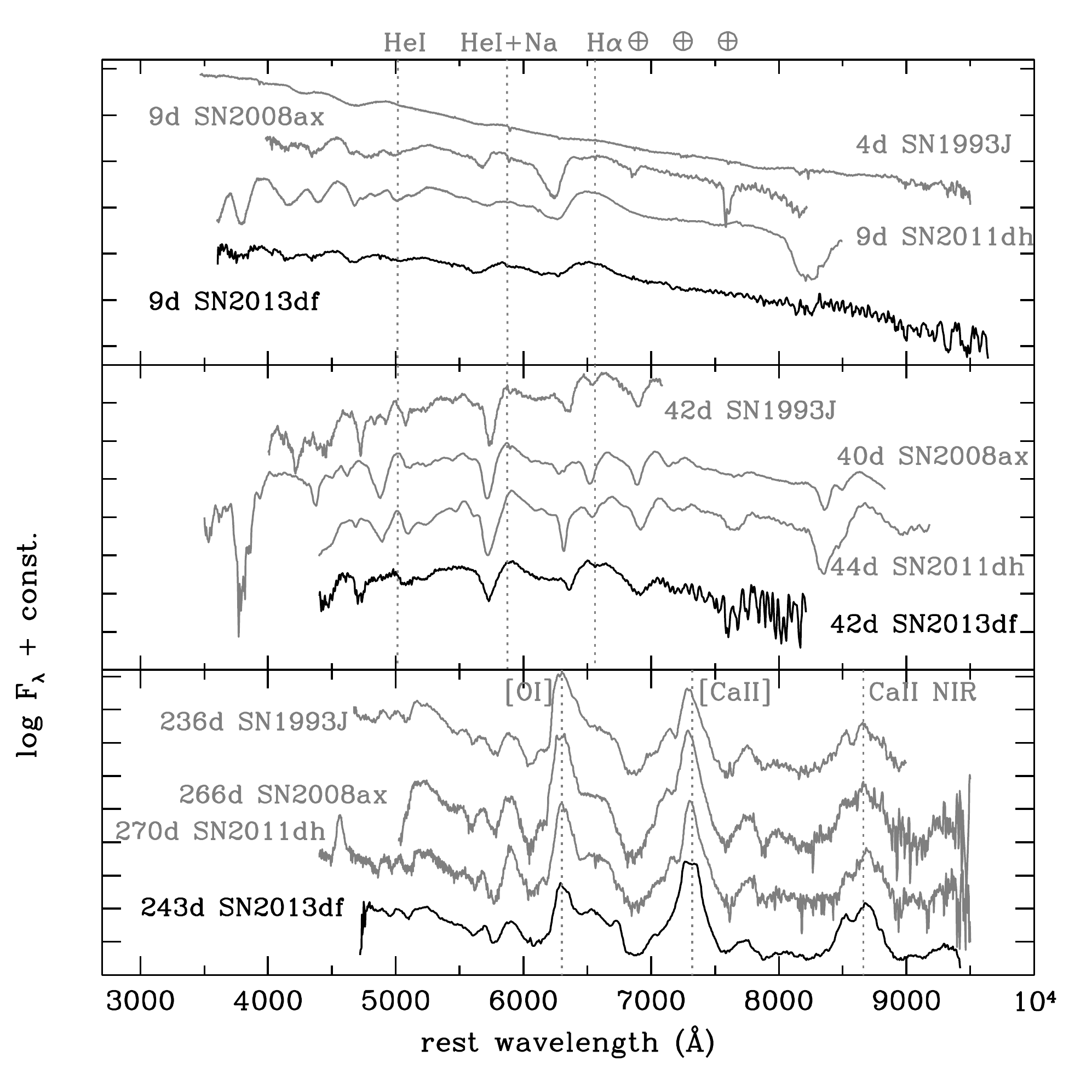}
 \caption{Comparison of early (4-9 days since explosion) intermediate (40-44 days after explosion) and late (236-270 days after explosion) spectra of SN~2013df with those of type IIb SNe 1993J, 2008ax and 2011dh. The original sources of these data are \citealt{Barbon95} for the spectra at 4 and 42 days of SN~1993J, and an unpublished spectrum taken at the 1.22m Galileo Telescope in Asiago-Italy at 236 days. For SN~2008ax, the spectrum at 9 days is from \citealt{pasto08ax} and the spectra at 40 and 266 days from \citealt{tau08ax}. For SN~2011dh, the spectrum at 10 days is from \citealt{Arcavi11dh}, the one at 44 days is from \citealt{Ergon11dh100D} and the one at 270 days from \citealt{shivvers2013}. All spectra have been redshift corrected and dereddened assuming the values given in Table~\ref{abs-Vmag}.}
 \label{fig:specearlylate}
\end{figure}

In Figure \ref{fig:specearlylate} we present a comparison of coeval spectra of SNe 2013df, 1993J, 2008ax and 2011dh. We have obtained the comparison spectra from the WISEREP\footnote[15]{http://www.weizmann.ac.il/astrophysics/wiserep/} \citep{yaron12} database. All the spectra have been dereddened  and corrected for their host galaxy recession velocities assuming the reddenings and redshifts given in Table~\ref{abs-Vmag}. At an early phase, the spectrum of SN~2013df is quite blue but shows stronger Balmer lines and  He\,{\sc i} 
$\lambda$5876 than the even bluer and almost featureless spectrum of SN~1993J. In contrast, the lines in SN~2013df's early spectrum are not as intense as in SNe~2008ax and 2011dh at a coeval phase.\\
In the middle panel of Figure \ref{fig:specearlylate}, we compare the above mentioned SNe at a phase of about 40 days. SN~2013df shows similarities to all three comparison SNe but is most similar to SN~1993J. Specifically, at this phase both SNe 2013df and 1993J have more intense H$\alpha$ absorption than SN~2008ax, but less prominent than SN~2011dh. The He\,{\sc i} $\lambda$$6678$ in the red wing of the H$\alpha$ emission is less prominent in SN~2013df than in the comparison SNe at this phase.\\
In the bottom panel of Figure \ref{fig:specearlylate} we have depicted nebular spectra for the above mentioned SNe IIb. All spectra are very similar, but it is noteworthy that [O\,{\sc i}] is less prominent in SN~2013df than in the rest of the cases, and its profile is more similar to that of SN~2011dh than the ones of SNe~1993J and 2008ax. The [Ca\,{\sc ii}] line is more prominent and boxy in SN~2013df than in the rest of the SNe. Although many effects such as mixing can affect the [Ca\,{\sc ii}]/[O\,{\sc i}] flux ratio in SN nebular spectra, it is expected to be sensitive to the progenitor's initial mass, increasing with decreasing main sequence mass \citep{franchev87,franchev89}. We note that this ratio for SN~2013df, compared to that of for example SN~1993J, could lead to the unrefined claim of SN~2013df's progenitor being less massive than SN~1993J. Interestingly this is in accordance with the estimation for SN~2013df's progenitor initial mass done above via the [O\,{\sc i}] $\lambda\lambda$$6300$, 
$6364$ luminosity measurements. We also note that the features in the nebular spectrum of SN~2013df at 6500-6600~\AA{} and 6700-6800~\AA{} discussed in the previous section, which we think are components of H$\alpha$, are not seen in the comparison SNe at coeval phase, specially the 6700-6800~\AA{} feature. 

\section{Constraints on the progenitor's properties}
\label{subsection:constraints}

Optical observations of CCSNe soon after shock breakout may help constrain progenitor characteristics. Clues can be given by early photometry (one day after explosion) as described in \cite{chefran08} and \cite{NkSari10}, thanks to colour temperature derived from early spectra \citep{RbWx2011} and by flash spectroscopy, as described by \cite{Gal-Yam14}.\\
 Recently, \cite{NkPiro14} studied the conditions for double-peaked SN light curves. They found that non standard progenitors formed by a compact massive core surrounded by an extended low-mass envelope can reproduce the main features of double-peaked light curves. They also derived a series of analytic formulae to constrain the mass of the extended material $ M_{\rm ext}$, its radius $ R_{\rm ext}$, and the radius of the core $ R_{\rm core}$:

\begin{itemize}
 \item From the time of the first peak in the LC ($t_{\rm p}$), the extended material mass prior to explosion is given by:

\begin{equation}
M_{\rm ext}\approx 8\times10^{-3}  E_{51}^{0.43} \kappa_{0.34}^{-0.87} \left(\frac{ M_{\rm core}}{3 \rm M_{\odot}}\right)^{-0.3}\left(\frac{ t_{\rm p}}{{\rm 1d}}\right)^{1.75} \rm M_{\odot}
\label{eq:mext}
\end{equation}

 \noindent  where $ M_{\rm ext}$ is taken to be the mass between the radii $R_{\rm ext}$/3 and $R_{\rm ext}$. $ E_{51}$ is the kinetic energy $ E/10^{51}$erg,  $ \kappa_{0.34}$ is the opacity $ \kappa / 0.34$ cm$^{2}$~s$^{-1}$, $ M_{\rm core}$ is the ejected core mass (not including mass of the compact remnant), and $ t_{\rm p}$ is the time in days at which the first peak in the LC takes place.\\

\item From the bolometric luminosity at the first peak ($L$), the radius of the extended material can be constrained by means of:

 \begin{equation}
 R_{\rm ext}\approx 10^{13} \kappa_{0.34}^{0.74} E_{51}^{-0.87}  L_{43}\left(\frac{M_{\rm core}}{3 \rm M_{\odot}}\right)^{0.61}\left(\frac{ t_{\rm p}}{\rm{1d}}\right)^{0.51} {\rm cm}
\label{eq:rext}
 \end{equation}
\noindent where, $ L_{43}$ is $\frac{ L}{10^{43}\rm erg\,s^{-1}}$ \\

\item From the luminosity at the minimum after the first peak ($L_{\rm min}$) an upper limit to the core radius may be derived:

\begin{equation}
  R_{\rm core}\lesssim2.5\times 10^{11} \kappa_{0.2}^{0.9} E_{51}^{-1.1}\left(\frac{ L_{\rm min}}{10^{41}{\rm erg\:s^{-1}}}\right)^{1.3}\left(\frac{ M_{\rm core}}{3 \rm M_{\odot}}\right)^{0.85}{\rm cm}
\label{eq:rcore}
\end{equation}

\noindent  where $\kappa_{0.2}$ is the opacity $\kappa / 0.2$ cm$^{2}$~s$^{-1}$.

\end{itemize}

In order to derive some characteristics of SN~2013df's progenitor, we made use of the following considerations:

\begin{enumerate}

\item We assumed two different constant optical opacities depending on the stage of the SN evolution. At first (for Formulae \ref{eq:mext} and \ref{eq:rext}) the LC is powered by the low mass extended hydrogen gas. As in \cite{NkSari10}, for a hydrogen envelope with cosmic abundances we considered $\kappa=0.34$ cm$^{2}$~g$^{-1}$. The second phase of the LC is powered by the compact hydrogen deficient core (Formula \ref{eq:rcore}), for which we considered  $\kappa=0.1$ cm$^{2}$~g$^{-1}$, the same value as the one we used in the modelling of the bolometric LC.

\item As noted in \cite{NkPiro14}, estimates for $R_{\rm ext}$ can be obtained even if only the luminosity in one band is available, though less accurately. In our case the first maximum of the light curve was only detected in the \textit{R} band, so we use $ L$ and $ t_{\rm p}$ as derived from that band and thus estimate only a lower limit to $ R_{\rm ext}$. 

 \item We assumed the initial mass range for SN~2013df's progenitor to be that given by \cite{VD13c}: $13-17$~$ \rm M_{\odot}$. We have estimated a range for the ejected core mass of $2-3.6$ M$_{\odot}$. To do this we considered the relation between the He-core mass and main sequence mass for the stellar evolution of a single star given by \cite{sugimoto80}, and assumed that the remnant mass is $\sim 1.4$ $\rm M_{\odot}$. Although the core mass we have estimated here is for a single star and the progenitor of SN~2013df could potentially form part of a binary system, we don't expect the core mass to be affected very much. For example, in the case of SN~2011dh, the core masses derived for its progenitor by \cite{Melina12} resulted similarly to that obtained by \cite{Benvenuto13},  considering SN~2011dh's progenitor evolved in a binary system.

\end{enumerate}

 We characterised SN~2013df's progenitor as summarized in Table~\ref{constraints}.

\begin{table*}
\centering
\caption{Estimates for the extended mass ($M_{\rm ext}$), radius ($ R_{\rm ext}$) and core radius ($ R_{\rm core}$) of SN~2013df's progenitor, calculated following the procedure described in \protect\cite{NkPiro14} for ``non standard progenitors''. For comparison, the values obtained for SNe 1993J and  2011dh in \protect\cite{NkPiro14} are also presented.}
\begin{tabular}{ccccccccc} \hline 
  SN  &$t_{\rm p}$ & $ L$& $ L_{\rm min}$ & $ E_{51}$   &    $ M_{\rm core}$   & $ M_{\rm ext}$  & $ R_{\rm ext}/R_{\odot}$ & $R_{\rm core}$  \\ 
    &(d) & (erg s$^{-1}$)&(erg s$^{-1}$)& (erg)      &  ($M_{\odot}$)       &  ($M_{\odot}$)     &  & ($10^{11}$ cm) \\
  \hline
2013df&$3.5$& L$_{R}=\; $$2.5\times10^{42}$ & $2\times10^{42}$ &$0.4-1.2$  &$2-3.6$ &  $0.05-0.09$     & $64-169$ & $<45-238$ \\ 
1993J&$3$&L$_{\rm bol}=1\times10^{43}$&$4\times10^{41}$&$1.31$&$2.23$&$0.06$&$288$ &$<9$ \\
2011dh&$0.27-0.85$&$-16.8\leq M_{g}\leq -15.5$& $2-2.5\times10^{41}$&$0.6-1$&$2.5$&$0.0007-0.006$&$288-503$&$<5-7$\\

\hline

\end{tabular} 
\label{constraints}
\end{table*}

Our result for SN~2013df's $ M_{\rm ext}$ is of the order of that obtained for SN~1993J in \cite{NkPiro14}, and an order of magnitude above the ones derived for SN~2011dh. As noted in \cite{NkPiro14}, $M_{\rm ext}$ is not the total mass of the envelope surrounding the core, but just the fraction distributed around $ R_{\rm ext}$. The similarity between $M_{\rm ext}$ for SNe 2013df and 1993J suggests that the mass of the hydrogen shell in SN~2013df's progenitor prior to explosion is similar to that of SN~1993J, i.e. $ M_{\rm H}\sim0.2$ $\rm M_{\odot}$ \citep{Ws94}, and slightly above the value for SN~2011dh $ M_{\rm H}\sim0.1$ $\rm M_{\odot}$ \citep{Melina12}. Concerning the extended radius of SN~2013df, we have derived a range of $64-169$ $ R_{\odot}$. We stress that this range is based on the \textit{R} luminosity instead of bolometric luminosity, which makes it a lower limit, as $ L_{ R}< L_{\rm bol}$. Taking into account the spectral and photometric similarity between SNe 
2013df and 1993J, SN~2013df's bolometric 
luminosity at first peak could be about a factor of 2 times brighter than the \textit{R} band specific luminosity. This could increase the estimate for  SN2013df's $R_{\rm ext}$ by a factor of 1.5. However, we must also take into account that for SN~1993J, $R_{\rm ext}$ derived by \cite{NkPiro14} is a factor 1.9 below that obtained from the modelling of the light curve, and a similar underestimate could also take place for SN~2013df. Multiplying our lower limit range on the extended radius of SN~2013df's progenitor by these two factors, brings it closer to the value derived by \cite{VD13c} from the pre-SN HST archival images, i.e. $545\pm65$$ R_{\odot}$. Finally we note that the range we have obtained for the upper limit to SN~2013df's core radius is much larger than the ones estimated for both SNe 1993J and 2011dh. Given the similar ejecta masses (Table~\ref{abs-Vmag}) this could suggest that the progenitor core of SN~2013df was not as compact as for SNe 1993J and 2011dh.

\section{Conclusions}

We have presented the analysis of our optical observations for the type IIb SN~2013df (complemented by UV data taken by \textit{SWIFT}) spanning from a few days up to 250 days after explosion. We have found SN~2013df to share common characteristics with previous type IIb SNe. Photometrically, similar to SNe 1993J and 2011fu,  SN~2013df's LCs present two peaks in all bands. The \textit{R} absolute magnitude LC evolves from a first maximum of $-18.12$ to a minimum value of $-17.20$, and then increases its brightness until reaching a secondary maximum of $-17.72$  approximately twelve days after the minimum. In common with SNe 1993J, 2008ax and 2011dh, the late decline rates are faster than the radioactive decay of $^{56}$Co.\\  
From the modelling of the bolometric light curve (not including the first peak), we have obtained an explosion energy of $0.4-1.2 \times 10^{51}$erg, a $^{56}$Ni mass in the range $0.1-0.13$ M$_{\odot}$ and a total ejected mass of $0.8-1.4$ M$_{\odot}$. These results are overall consistent with other modelled IIb SNe.\\
The earliest spectrum of our sequence shows a flat-bottomed  H$\alpha$ absorption indicative of the presence of an extended H shell in the ejecta, which is the outer layer of the progenitor's envelope.
The presence of conspicuous hydrogen in our spectra up to later phases than in SN~2008ax, is indicative of possibly a slightly larger mass of the H shell in SN~2013df. Furthermore, the resemblance of SN~2013df's spectra with those of SN~1993J indicate that the H shell of both SNe could be similar ($ M_{\rm H}\sim0.2$ $\rm M_{\odot}$) and thus possibly above the value estimated for SN~2011dh ($ M_{\rm H}\sim0.1$ $\rm M_{\odot}$ ).
Our nebular spectra are characterised by the presence of two components around 6500-6600~\AA{} and 6700-6800~\AA{} that are possibly  H$\alpha$ in emission caused by asymmetrical interaction of the ejecta with circumstellar material. Further spectral modelling should be done in order to confirm this speculative claim. From the [O\,{\sc i}] $\lambda$$\lambda$$6300$, $6364$ luminosities we have derived a rough estimate to the initial mass of SN~2013df's progenitor $\sim 12-13 \rm M_{\odot}$. From the analysis of the nebular profiles of [O\,{\sc i}] and [Ca\,{\sc ii}], we conclude that the substructure they exhibit is indicative of the presence of different clumps where these species are being excited in the ejecta. 
\\
Finally, we have followed the procedure described in \cite{NkPiro14} to add some additional constrains on some progenitor characteristics of SN~2013df assuming that it was a non-standard progenitor formed by a compact core and an extended low-mass envelope. We have estimated a mass range for the extended material similar to SN~1993J, which in accordance to our spectra, leads us to believe both SNe 1993J and 2013df had similar hydrogen shells prior to explosion. In addition, we estimated a lower limit to the radius for SN~2013df's progenitor of $64-169$ $R_{\odot}$. At last, we have obtained an upper limit to the core radius of SN~2013df's progenitor $ R_{\rm core}<( 45-238) \times 10^{11}$ cm which are well above those derived for SNe 1993J and 2011dh.\\
SN~2013df is the third type IIb SN presenting double-peaked light curves in all its optical bands. In addition, its progenitor has been identified in archival images. Further modelling of the SN data and future observations of the SN field may help extract more information on the possible yellow supergiant that exploded as SN~2013df and the conditions that produced its observed properties.

\section*{Acknowledgements}

We would like to thank the anonymous referee for his/her comments, which helped to greatly improve this manuscript.
A.M.G. acknowledges financial support by the Spanish Ministry of Science and Innovation (MICINN) grant AYA2011-24704/ESP, by the  ESF EUROCORES Program EuroGENESIS (MINECO grants 
EUI2009-04170), SGR grants of the \textit{Generalitat de Catalunya} and by the EU-FEDER funds. N.E.R. acknowledges the support from the European Union Seventh Framework Programme (FP7/2007-2013) under grant agreement n. 267251 ``Astronomy Fellowships in Italy'' (AstroFIt). S.B., E.C., A.P., L.T. and L.T. are partially supported by the PRIN-INAF 2011 with the project ”Transient Universe: from ESO Large to PESSTO”. S.T. acknowledges support by the Transregional Collaborative Research Centre TRR 33 of the German Research Foundation (DFG).

 We are very thankful to Anders Jerkstrand for providing us helpful information for the analysis of the nebular spectra and its implications on the estimation of the progenitor initial mass.

We would also like to thank F. Ciabattari from the ISSP for providing us the discovery and confirmation image of the SN taken at the \textit{Osservatorio di Monte Agliale} (Lucca, Italy). We also are grateful to S. Donati and K. Itagaki for providing us early time images of the SN taken by them at San Vito (Lucca, Italy) and the Koichi Astronomical Observatory (Japan) respectively.

This work is partially based on observations made with the \textit{Telescopi Joan Or\'o} of the Montsec Astronomical Observatory, which is owned by the \textit{Generalitat de Catalunya} and operated by the Institute for Space Studies of Catalunya (IEEC); the Liverpool Telescope which is operated by Liverpool John Moores University with financial support from the UK Science and Technology Facilities Council, the 10.4-m \textit{Gran Telescopio de Canarias}, and the \textit{Telescopio Nazionale Galileo} operated by INAF (\textit{Instituto Nazionale dia Astorfisica}) in the Spanish \textit{Observatorio del Roque de los Muchachos} (ORM) of the \textit{Instituto de Astrof\'isica de Canarias}; the 1.82-m Copernico Telescope operated by INAF - \textit{Osservatorio Astronomico di Padova} and the 1.22-m Galileo Telescope of \textit{Dipartimetno di Fisica e Astonomia (Universit\'a di Padova)} at the Asiago Observatory (Italy). This research has made use of the NASA/IPAC Extragalactic Database (NED) which is operated by 
the Jet Propulsion 
Laboratory, California Institute of Technology, under contract with the National Aeronautics and Space Administration.\\

\addcontentsline{toc}{chapter}{Bibliography}
\markboth{Bibliography}{Bibliography}
\bibliographystyle{mn2e}

\appendix
\section{Instrumental Set-ups}
\label{appendix}
\begin{enumerate}

\item Ground-based photometry
 
    \begin{enumerate}
    \item \textit{UBVRI} with MEIA (field of view of 12'.3 x 12'.3, pixel scale of $0{\farcs}13$ pix$^{-1}$) at the \textit{Telescopi Joan Or\'o} (TJO; 0.82m) of the \textit {Observatori Astron\'omic del Montsec} in Catalunya (OAdM, Spain).
    \item \textit{UBVRI} with AFOSC (field of view of 8'.1 x 8'.1, pixel scale of $0{\farcs}473$ pix$^{-1}$) at the Copernico Telescope (1.82m) of the Asiago Observatory (Italy).
   \item \textit{uBVri} with RATCAM (field of view of 4'.6 x 4'.6, pixel scale of $0{\farcs}135$ pix$^{-1}$) at the Liverpool Telescope (LT; 2.0m) of the Roque de los Muchachos Observatory (ORM; Spain).
     \item \textit{uBVri} with IO:O (field of view of 10' x 10', pixel scale of $0{\farcs}30$ pix$^{-1}$) at the LT.
     \item \textit{UBVRI} with LRS (field of view of 8'.6 x 8'.6  with a pixel scale of $0{\farcs}252$ pix$^{-1}$) at the \textit{Telescopio Nazionale de Galileo} (TNG; 3.58m) of ORM.

     \item Unfiltered image provided by Fabrizio Ciabattari from the ISSP taken with a FLI Proline CCD (field of view of 20'.4 x 19'.8, pixel scale of $2{\farcs}32$ pix$^{-1}$) at the Newtonian Telescope (0.5m) of the \textit{Osservatorio di Monte Agliale}\footnote[16]{http://www.oama.it/} (Lucca, Italy) .

      \item Unfiltered image provided by Sauro Donati taken with a SBIG ST10 dual camera (field of view of 12'.4 x 18'.4, pixel scale of $1{\farcs}52$ pix$^{-1}$) with a  Schmidt-Cassegrain Telescope (0.3m) at San Vito (Lucca, Italy).

      \item Unfiltered image provided by Koichi Itagaki\footnote[17]{http://www.k-itagaki.jp/} taken with a Bitran BT-214E CCD (KAF 1001E) (field of view of 28'.2 x 28'.2, pixel scale of $1{\farcs}65$ pix$^{-1}$) with a reflector telescope (0.5m at f/6) of the Itagaki Astronomical Observatory (Japan).

      \item Unfiltered image provided by Stan Howerton\footnote[18]{http://www.itelescope.net/} taken with a KAF-6303 CCD (field of view of 37'.41 x 24'.94, pixel scale of $0{\farcs}73$ pix$^{-1}$) at the iTelescope.net T18 (0.32m) (Spain). 
      
      \item Images provided by Norbert Schramm\footnote[19]{http://njstargazer.org/} taken with a Orion Star Shoot Pro V2 one shot camera (field of view of 66.5' x 100', pixel scale of $1{\farcs}98$ pix$^{-1}$) at a Schmidt/Newtonian telescope (0.20m) (Oxford, U.S.A.).

     \item \textit{V} images provided by Stan Howerton taken with a SBIG ST-10XME CCD camera (field of view of 40'.4 x 60', pixel scale of $1{\farcs}65$ pix$^{-1}$) at the iTelescope.net T5 (0.25m) (New Mexico, USA).

         \end{enumerate}

\item Space-based photometry\\

  \textit{UBV} and ultraviolet \textit{UVW2}, \textit{UVM2}, and \textit{UVW1} data obtained with the 30-cm modified Ritchey-Chretien UV/optical telescope (UVOT) equipped with  a micro channel plate intensified CCD (MIC), on board the \textit{SWIFT} satellite.\\

   \item Spectroscopy
  \begin{enumerate}
   \item Grism 4 (spectral range = 3500-8450 \AA; resolving power = 613), holographic grism VPH6 (range = 4500-10000 \AA; resolving power = 500) with AFOSC at the Copernico telescope of the Asiago Observatory (Italy).
   \item  300 lines mm$^{-1}$ grating (spectral range = 3350-7850 \AA; resolving power = 700) with B\&C at the 1.22m Galileo telescope at the Asiago Observatory (Italy).
   \item Grism R500R (spectral range = 4800-10000 \AA; resolving power = 587) with Osiris at the 10.4m \textit{Gran Telescopio de Canarias} at the \textit{ORM} (Spain).
  \end{enumerate}
\end{enumerate}

\label{lastpage}
\end{document}